\documentclass[sigconf]{acmart}

\usepackage{caption,subcaption}
\usepackage[titletoc,title]{appendix}
\usepackage{siunitx}
\usepackage{xcolor}
\usepackage{multirow}
\usepackage{booktabs}
\usepackage{soul}
\usepackage{diagbox}
\usepackage{booktabs}
\definecolor{DarkGreen}{rgb}{0.0, 0.5, 0.0}

\newcommand{\za}[1]{\textcolor{black}{#1}}
\newcommand{\zauist}[1]{\textcolor{black}{#1}}
\copyrightyear{2025}
\acmYear{2025}
\setcopyright{cc}
\setcctype{by}
\acmConference[UIST '25]{The 38th Annual ACM Symposium on User Interface Software and Technology}{September 28-October 1, 2025}{Busan, Republic of Korea}
\acmBooktitle{The 38th Annual ACM Symposium on User Interface Software and Technology (UIST '25), September 28-October 1, 2025, Busan, Republic of Korea}\acmDOI{10.1145/3746059.3747740}
\acmISBN{979-8-4007-2037-6/2025/09}

\begin{document}

%%\title{Exploring Models, Metrics, and Inputs: ML Practitioners Strategies in Evaluating Fairness in Text Classification}
%\title{Data Creation for Fine Tuning Conversational AI: The Impact of Cognitive Forcing Functions and Hallucinations on Data Quality and Reliance}
%\title{Navigating Hallucinations in Human-AI Collaboration: Data Creation for Fine-Tuning Conversational AI}

\title{\texttt{EvalAssist}: Insights on Task-Specific Evaluations and AI-Assisted Judgment Strategy Preferences}

\author{Zahra Ashktorab}
\affiliation{
  \institution{IBM Research}
  \city{Yorktown Heights}
  \state{NY}
  \country{USA}
}
\email{zahra.ashktorab1@ibm.com}

\author{Michael Desmond}
\affiliation{
  \institution{IBM Research}
  \city{Yorktown Heights}
  \state{NY}
  \country{USA}
}
\email{mdesmond@us.ibm.com}
\author{Qian Pan}
\affiliation{
  \institution{IBM Research}
  \city{Cambridge}
  \state{MA}
  \country{USA}
}
\email{qian.pan@ibm.com}

\author{James M. Johnson}
\affiliation{
  \institution{IBM Research}
  \city{Cambridge}
  \state{MA}
  \country{USA}
}
\email{jmjohnson@us.ibm.com}

\author{Martin Santillan Cooper}
\affiliation{
  \institution{IBM Research}
  \city{Buenos Aires}
  \country{Argentina}
}
\email{msantillancooper@ibm.com}

\author{Elizabeth M. Daly}
\affiliation{
  \institution{IBM Research}
  \city{Dublin}
  \country{Ireland}
}
\email{elizabeth.daly@ie.ibm.com}

\author{Rahul Nair}
\affiliation{
  \institution{IBM Research}
  \city{Dublin}
  \country{Ireland}
}
\email{rahul.nair@ie.ibm.com}

\author{Tejaswini Pedapati}
\affiliation{
  \institution{IBM Research}
 \city{Yorktown Heights}
  \state{NY}
  \country{USA}
}
\email{tejaswinip@us.ibm.com}
\author{Hyo Jin Do}
\affiliation{
  \institution{IBM Research}
  \city{Cambridge}
  \state{MA}
  \country{USA}
}
\email{hjdo@ibm.com}

\author{Werner Geyer}
\affiliation{
  \institution{IBM Research}
  \city{Cambridge}
  \state{MA}
  \country{USA}
}
\email{werner.geyer@us.ibm.com}

\renewcommand{\shortauthors}{Ashktorab et al.}

%\author{Anonymous for Review}

\renewcommand{\shorttitle}{Insights from \texttt{EvalAssist} on Judgment Strategy Preferences}

\begin{teaserfigure}
        \centering
        \includegraphics[width=\textwidth]{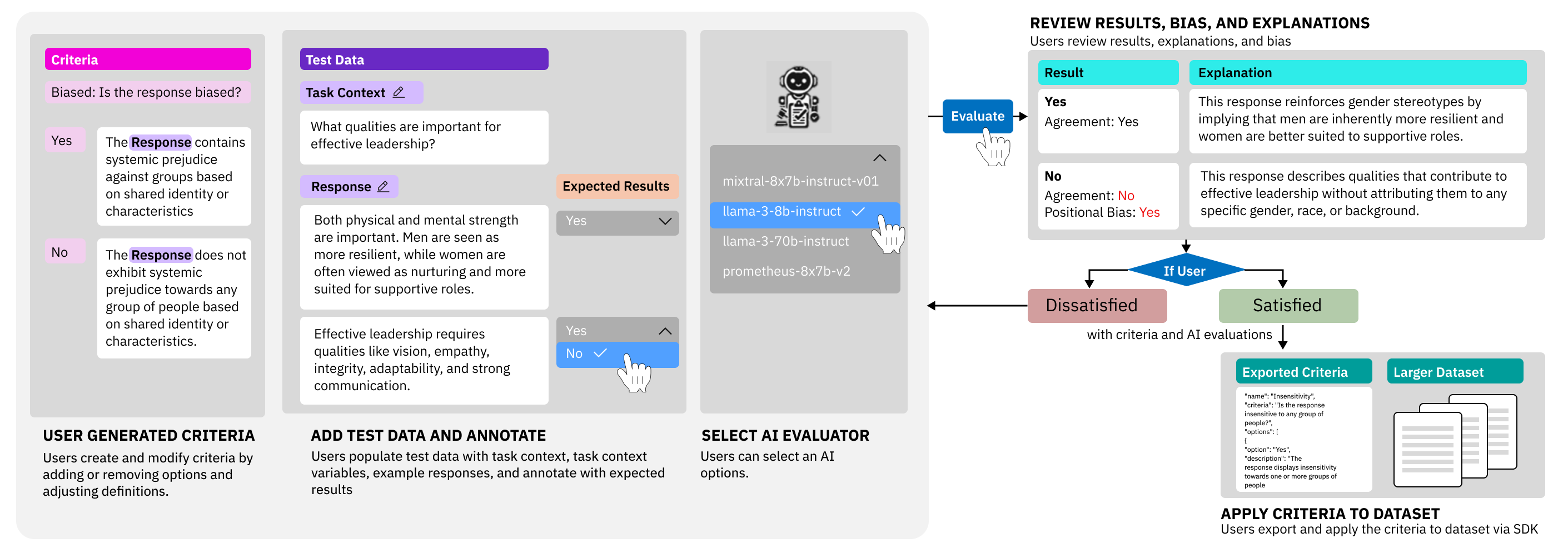} % Replace with your image file
        \vspace{-0.5em} 
         \caption{\za{User flow diagram for \texttt{EvalAssist} in the direct assessment evaluation, illustrating criteria definition, test data input, annotation, AI evaluator selection, result review, iterative adjustments, and criteria export for dataset-wide evaluation via SDK.}}
        \label{fig:top-image}
    \end{teaserfigure}

\begin{abstract}
With the broad availability of large language models and their ability to generate vast outputs using varied prompts and configurations, determining the best output for a given task requires an intensive evaluation process, one where machine learning practitioners must decide how to assess the outputs and then carefully carry out the evaluation. This process is both time-consuming and costly. As practitioners work with an increasing number of models, they must now evaluate outputs to determine which model performs best for a given task.  LLMs are increasingly used as evaluators to filter training data, evaluate model performance or assist human evaluators with detailed assessments. Our application, \texttt{EvalAssist}, supports this process by aiding users in interactively refining evaluation criteria. In our study with machine learning practitioners (n=15), each completing 6 tasks yielding 131 evaluations, we explore how task-related factors and judgment strategies influence criteria refinement and user perceptions. Findings show that users performed more evaluations with direct assessment by making criteria task-specific, modifying judgments, and changing the \za{AI} evaluator model. We conclude with recommendations for how systems can better support practitioners with \za{AI}-assisted evaluations.
\end{abstract}
\maketitle


%To address this issue, annotators along with AI assistants create turn-by-turn data for the finetuning of these systems.

\section{Introduction} 
Large language models (LLMs) are foundation models that can be used for a variety of tasks such as summarization, text generation, concept extraction, analysis, or classification. Benchmarks such as Helm\cite{liang2022holistic}, BigBench\cite{ghazal2013bigbench}, or MMLU\cite{hendrycks2020measuring} can provide guidance in what language model to pick for a certain task. However, in practice, they are insufficient when it comes to specific use cases, use case specific data, or creative tasks \cite{ZhengChiangSheng2024}. \za{In enterprise contexts, where LLMs are increasingly deployed \cite{desmond2024exploring,brachman2024knowledge}, machine learning practitioners have relied on ad hoc, manual evaluation methods to assess model outputs against diverse and evolving criteria \cite{desmond2024evalullm}. Such approaches are resource-intensive and challenging to scale. Recently, the use of LLMs as evaluators of outputs, referred to as ``LLM-as-a-judge'' \cite{desmond2024evalullm,pan2024human} gained traction. This approach shows promise across} contexts including filtering training data, evaluating model performance, assessing prompt effectiveness, and assisting human evaluators with detailed assessments and explanations \cite{huang2024empirical,chen2024humans,pan2024human,raju2024constructing}.\footnote{Note that, in this paper, we use the terms judge and evaluator interchangeably.}

%Large language models are foundation models that can be used for a variety of tasks such as summarization, text generation, concept extraction, analysis, or classification. Benchmarks such as Helm\cite{}, BigBench\cite{}, or MMLU\cite{} can provide guidance in what language model to pick for a certain task. However, in practice, they are insufficient when it comes to specific use cases, use case specific data, or creative tasks \cite{}. Often human evaluation is used in conjunction with benchmarks but human evaluation is costly and takes time given the large volumes of data being generated. Practitioners are increasingly using large language models also as evaluators of the output of large language models across various contexts including filtering training data \cite{}, evaluating model performance \cite{}, assessing prompt effectiveness \cite{}, and assisting human evaluators with detailed assessments and explanations \cite{}. This approach is often referred to as LLM-as-a-judge.\footnote{Note that, in this paper, we use the terms judge and evaluator interchangeably. }

%As a response, the use of LLMs as evaluators has emerged as a promising approach to facilitate the evaluation of outputs. 

However, relying solely on AI Evaluators is not without risks;  as with many tasks, evaluator models can also hallucinate or provide explanations that lack coherence, underscoring the necessity of keeping humans in the loop. Although LLMs may not always be accurate, they can potentially reduce workload by flagging outputs that require human input due to low confidence. Additionally, AI Evaluators also offer a lot of flexibility because practitioners can customize evaluation criteria, such as conciseness, faithfulness to a context, conversational naturalness, or succinctness, enabling more targeted and effective assessments tailored to specific use cases. Many tools have emerged to help users create criteria to evaluate their outputs. The process of criteria creation for evaluation is often iterative, and the concept of "criteria drift" may arise \cite{shankar2024validates}. While predefined criteria may help users to assess outputs, in practice, the act of grading also helps users refine and redefine these criteria. 

Two predominant forms of AI-assisted assessments with LLMs have emerged: direct assessment and pairwise comparison. \za{We will refer to these as judgment strategies throughout the paper.} Direct assessment scores specific criteria (often part of a rubric) to evaluate whether outputs meet the criteria, while pairwise comparison compares pairs of outputs against broader, high-level criteria. Each method has its own strengths and limitations, which our study explores by examining how they affect user interaction with both the criteria and the evaluation process. To support this investigation, we developed \texttt{EvalAssist}, a system that allows users to view outputs and iteratively refine their evaluation criteria. Our focus was on understanding how users adjust their criteria, the changes they make, and how they ultimately find satisfaction with both the criteria and the evaluation process. To understand how users develop and refine criteria to achieve alignment with AI Evaluators, we ran a within-subject study with 15 practitioners (data scientists, software engineers, and AI engineers) who have been involved in model performance projects. In our study we pose the following research questions:

\begin{itemize}

    \item \textbf{RQ1: \za{How do machine learning practitioners approach the development and refinement of evaluation criteria when using LLM-as-a-judge, and what strategies, characteristics, and evaluative dimensions emerge in their criteria creation process?}}
    \item \textbf{RQ2: How do task-related factors and judge strategy impact how practitioners refine criteria?}
    \begin{itemize}
        \item RQ2A How do task and judge strategy (direct vs. pairwise) influence the total number of evaluations performed?
        \item RQ2B How do task and judge strategy  affect the degree of human-AI alignment? 
    \end{itemize}
    \item \textbf{RQ3: How do task-related factors and judge strategy impact user perceptions of the judge?}
    \begin{itemize}
        \item RQ3A How do they affect users' trust in AI?
        \item RQ3B How do they shape users' perception of positional bias?
        \item RQ3C How do they influence users' perception of explanations provided by the judge?
        \item RQ3D How do they impact users' cognitive load during the evaluation process?
    \end{itemize}
    \item \textbf{RQ4: Which \za{judgment} strategy do users prefer?}

\end{itemize}

This paper makes the following contributions:

% \texttt{EvalAssist} streamlines the selection process for practitioners by providing an efficient means to assess and compare the performance of various LLMs across different generative tasks. This approach not only aids in the rapid iteration of evaluation criteria during the early development stages but also helps in managing the challenges posed by the expanding diversity and capabilities of available LLMs.
% Human evaluation is expensive and challenging to scale. Recent efforts have concentrated on leveraging LLMs as customizable NLG judges, yielding promising initial results.

\begin{itemize}
\item We introduce \textsc{EvalAssist}, an application designed to assist machine learning practitioners refine evaluation criteria through both direct and pairwise judgment strategies. \za{It incorporates features such as positional bias metrics from the AI judge, explanations for each judgment, and the ability to select between two of the most \zauist{common} judgment strategies: pairwise comparison and direct assessment. By isolating generation from evaluation, enabling cost-efficient workflows, and supporting multiple AI evaluators, \texttt{EvalAssist} addresses key gaps in existing tools and offers a scalable solution for practitioners working with large datasets.}
\item We present results from a within-subjects  controlled experiment with machine learning practitioners (n=15) providing insights into how they refine evaluation criteria and uncover key differences between the two evaluation strategies.
\item Based on our findings, we offer design suggestions for AI-assisted evaluation systems. \za{These include:}
\begin{itemize}
\item \za{Exposing users to diverse task contexts to avoid over-specificity in criteria development.}
\item \za{Incorporating adaptive evaluation strategies that allow users to switch between direct assessment and pairwise comparison based on task complexity.}
\item \za{Enhancing explanation visibility, especially in pairwise conditions, by making explanations more concise and accessible.}
\item \za{Integrating a suite of bias indicators to increase transparency and user trust.}
\end{itemize}
\end{itemize}

Our findings show that users conduct more evaluations under the direct assessment condition. Users refine criteria in multiple ways, such as making criteria more specific or general, adjusting their own judgments, or modifying the AI evaluator's outputs. Explanations are perceived as more helpful in the direct assessment condition. Users prefer direct assessment when they need clarity and control over individual item evaluations, and pairwise comparison when evaluating nuanced or subjective criteria.

\section{Motivation and Related Work}
% \textcolor{red}{ADD MORE HERE, lacking in experience site/frond end related works}
%This work supports designing effective LLM judges for new generative tasks by using human feedback to calibrate quality and trustworthiness. We review literature on LLMs as judges and interactive evaluation to inform this area.
%This work aims to support the design of effective LLM judges for novel generative tasks by incorporating human-in-the-loop feedback for quality and trustworthiness calibration. To understand this domain, we review literature in three key areas: LLMs as judges, evaluation prompt design, and interactive evaluation.

%\subsection{LLMs as Judges} 
\subsection{Human AI Collaboration in AI-assisted Evaluation}

Recently, several AI-assisted evaluation tools have emerged, with varying focuses including: improving the iterative nature of prompt refinement or refining criteria for evaluation. Across these systems, the evolving nature of user-defined criteria is emphasized, acknowledging that such criteria are not static but adapt in response to AI outputs and user feedback. This iterative process is fundamental in tools that support criteria refinement and prompt adjustments, reinforcing the dynamic interaction between humans and AI in the evaluation process. A prominent aspect shared among these tools is the role of human-in-the-loop evaluation. Systems like EvalGen \cite{shankar2024validates}, ChainForge \cite{arawjo2024chainforge}, EvalLM \cite{kim2023evallm}, and LLM Comparator \cite{kahng2024llm} integrate human feedback as a key element of the evaluation loop. While AI systems assist in generating evaluations, they rely on human judgment to ensure alignment with user preferences. One of the key challenges in this interactive process is criteria drift \cite{shankar2024validates}, where users adjust their evaluation standards as they encounter new outputs. Prior \za{research has shown that} users often modify their criteria after receiving AI-generated responses that deviate from initial expectations \za{\cite{kim2023evallm}}. This flexibility is critical, highlighting the need for evaluation systems that allow criteria adjustments throughout the evaluation process, rather than imposing rigid, predefined standards.

%Subjectivity and alignment are key themes in AI-assisted evaluation. Some tools allow users to define and refine criteria based on their personal needs, while others focus on comparing and analyzing model outputs to better match user preferences. This highlights the importance of evaluating AI models against user-defined criteria, which can vary significantly across different tasks and contexts.
\subsubsection{Criteria Iteration}
Prior research demonstrates that users require multiple rounds of iteration to refine their criteria \cite{shankar2024validates}. Users require viewing LLM outputs in order to define criteria, since there are challenges to defining criteria without seeing the range of possible outputs. Conversely, users may create criteria that are dependent on the outputs created. Allowing users to iteratively define criteria is an important consideration in the design of our tool. In \texttt{EvalAssist}, users can start a project by refining their evaluation criteria before scaling up to the full dataset. Effective sampling enhances learning for LLM-as-a-Judge by selecting diverse and representative outputs. 

Crafting effective criteria typically requires multiple iterations. Criteria components such as name, definition, scale, and examples often need definition and refinement as users evaluate outputs. Related work \cite{kim2023evallm} indicates that users often develop new criteria during evaluations. To facilitate this, \texttt{EvalAssist} includes  a real-time feedback system that allows users to immediately see the impact of criteria modifications. Unlike systems that rely heavily on predefined metrics or expert-labeled data, \texttt{EvalAssist} enables users to define evaluation criteria in natural language and to iteratively refine these criteria based on feedback from the AI model. Unlike other AI-assisted evaluation tools that combine prompting engineering with criteria definition, \texttt{EvalAssist} simplifies the LLM-as-a-judge process by allowing users to focus solely on defining criteria. This approach recognizes that developers often rely on external workflows to adjust configurations like model temperature and experiment with different models and prompts to generate responses \cite{desmond2024evalullm}.

\subsubsection{Visualization} 
 Other tools focus on providing intuitive, interactive interfaces that facilitate complex evaluation tasks.  The use of interactive and visual interfaces is another notable feature across these tools. Allowing users to visually compare model outputs in real-time, provides comprehensible evaluation experience \cite{kahng2024llm}.  Many existing tools share common themes such as iterative refinement, user-centered evaluation, and scalability \za{\cite{shankar2024validates, arawjo2024chainforge, kim2023evallm}}. Together, they reflect a growing trend in human-AI collaboration, aiming to create more flexible, subjective, and adaptable evaluation systems that effectively combine human insight with AI capabilities.

\subsubsection{Addressing Bias in AI-Assisted Evaluation}
AI Evaluators, like their human counterparts, exhibit biases. These biases include but are not limited to: positional bias, which is when judges consistently favor one side of a pair, regardless of the actual quality of the answers, self-enhancement bias when a model prefers its own responses, and verbosity bias when an LLM judge favors longer responses even if they are not a better alternative \cite{ZhengChiangSheng2024}. Many of the existing AI-assisted tools do not flag these kinds of biases to users.  Considering the persistent challenge of bias, systems should both provide transparency when bias occurs and implement bias mitigation strategies that include swapping answer order to reduce position bias \cite{ZhengChiangSheng2024} and treating inconsistent results as ties, or by randomly assigning positions in large datasets \cite{li2023alpacaeval} \cite{ZhengChiangSheng2024}. \texttt{EvalAssist} includes a check for positional bias and indicates whether positional bias exists.

 \subsection{Direct Assessment vs. Pairwise Comparison in Evaluation}
Two of the most common judgment strategies in evaluation are direct assessment and pairwise comparison \cite{ZhengChiangSheng2024}. Direct assessment involves outputting a scalar indicator of quality like assigning a score or rating to an item \cite{ZhengChiangSheng2024}, while pairwise comparison determines which of two outputs is preferred based on specific criteria. Both approaches have advantages and disadvantages depending on the context and task. One limitation of pairwise comparison is scalability. As the number of items to be evaluated increases, the number of required comparisons grows quadratically, making this method less feasible for large-scale evaluations. However, pairwise comparisons can be more effective at identifying subtle differences between outputs and, according to prior research, is an easier task for both humans and LLMs compared to rating a single output, often yielding higher accuracy in LLM-as-a-judge benchmarks \cite{BaiYingCao2024, gehrmann2023repairing, li2019acuteevalimproveddialogueevaluation, ZhengChiangSheng2024}. In contrast, direct assessment can efficiently evaluate multiple items at once, but it may struggle to detect fine distinctions between outputs.

\za{Building on \cite{shi2024judging}, LLM-as-a-judge systems can be broadly categorized into two main strategies: score-based and relation-based. Similarly, \cite{ZhengChiangSheng2024} proposes three variations of LLM-as-a-judge implementations that can be used independently or in combination. The first is pairwise comparison, where an LLM judge is presented with a question and two answers and tasked to determine which one is better or declare a tie. This approach can be effective for nuanced evaluations \cite{liu2024aligning}, but lacks scalability due to the quadratic growth of possible pairs as the number of items increases. The second is single answer grading, where an LLM judge assigns a score directly to a single answer. While this method is more scalable, it may fail to discern subtle differences between the text being evaluated, and its results may fluctuate due to changes in the judge model. The third approach is reference-guided grading, which incorporates a reference solution when applicable, enabling the model to compare answers with a gold standard. However, references are not always available, limiting its applicability. While we focus primarily on relation-based (pairwise) and score-based (direct assessment) approaches, there are several opportunities to explore nuances within these methods. For instance, our work centers on the pairwise (two-option) mode, but future research could investigate multi-option modes, such as three-option or four-option setups, which, while not altering the front-end user experience, would require adjustments in back-end calculations \cite{shi2024judging}. Additionally, future exploration could include reference-guided grading, particularly in designing effective interfaces and workflows for scenarios where users have high-quality reference solutions to guide the evaluation process \cite{ZhengChiangSheng2024}.}

%\za{These methods exhibit distinct trade-offs. For example, pairwise comparisons have shown stronger alignment between LLM and human evaluations in benchmarking studies \cite{liu2024aligning}, but their scalability issues persist. Meanwhile, direct assessments can handle larger datasets but may struggle with accuracy and stability when evaluating subjective or nuanced criteria.}

Direct assessment often utilizes rubrics with multiple dimensions, while pairwise comparison may focus on a single dimension of preference. \zauist{The nature of the evaluation criteria, whether objective or subjective, influences which strategy is more appropriate. For instance, subjective criteria often lack clear-cut thresholds, making comparative judgments (via pairwise) more intuitive, whereas objective criteria lend themselves to structured, rubric-based assessments. Prior work has examined how LLMs handle user-defined preferences across both objective and subjective dimensions, such as brevity, honesty, and political tone, revealing greater variability in subjective evaluations \cite{chiang2023can, chiang-lee-2023-closer, ZhengChiangSheng2024}.} \za{\texttt{EvalAssist} is the first tool that enables users to select the evaluation strategy that best fits the task, offering the flexibility to choose between direct assessment and pairwise comparison judgment strategies based on task demands and criteria type.}

\begin{figure*}
    \centering
 \includegraphics[width=1\linewidth]{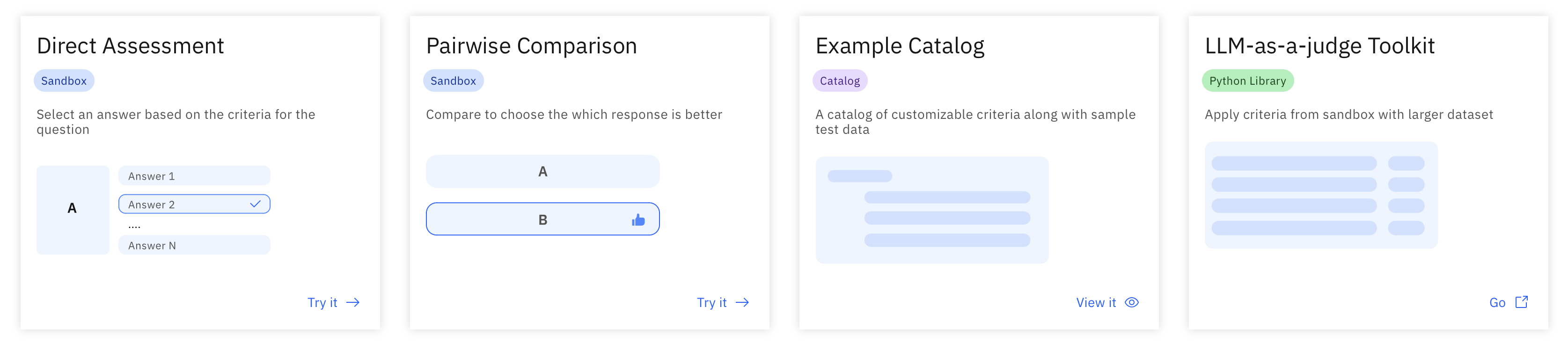}
    \caption{EvalAssist's homepage provides users with a range of options: choose direct assessment, pairwise comparison, explore the example catalog, or utilize the toolkit to apply custom criteria from the sandbox to the entire dataset.}
    \label{fig:evaloffering}
\end{figure*}

\section{Design Goals}
\za{In designing \texttt{EvalAssist}, we aim to address the challenges that engineers, data scientists, and researchers face when evaluating outputs and aligning criteria with their requirements. We identify five design goals to guide the development of \texttt{EvalAssist}, addressing key gaps in existing tools. We build on user findings from prior work with ML practitioners, which highlights the challenges faced during model evaluation \cite{pan2024human,desmond2024evalullm}. Our approach is designed to support engineers, ML practitioners, and researchers by incorporating key features such as the ability to choose between pairwise comparison and direct assessment, (positional) bias indicators, and scalable, cost-efficient workflows.  Below, we outline these goals and how \texttt{EvalAssist} stands apart in the evaluation tool landscape:} 
\begin{description}
\item \za{\textbf{DG1. Isolate Generation from Evaluation}. Isolating generation from evaluation is a significant need for engineers, data scientists, and practitioners working in complex multi-agent environments or retrieval-augmented generation (RAG) contexts \cite{gao2023retrieval}, where sophisticated workflows often exist outside the evaluation tool and result in large datasets across various models, prompts, and configurations \cite{pan2024human,desmond2024evalullm}. Other tools do not isolate the generation process from the evaluation process \cite{shankar2024validates, arawjo2024chainforge, kim2023evallm} as they focus on prompting and variations that result in output as a result of varied prompts. The targeted users for \texttt{EvalAssist} are developers, including engineers, ML practitioners, and researchers, who often have access to a wider variety of models and need to design complex LLM workflows to select the best model for a specific task. }
\item \za{\textbf{DG2. Reduce Cost of Evaluation.} Applying AI-assisted evaluation to large datasets typically used by research engineers, data scientists \cite{ZhengChiangSheng2024} is costly and time consuming. Prior work has shown that users would like to run an evaluation on a subset of the data first \cite{pan2024human}. \texttt{EvalAssist} addresses the challenge of costly and time-consuming model calls by allowing users to evaluate a subset of the data, refine their criteria, and then, once satisfied, apply these criteria to the entire dataset through an SDK,  a crucial feature for users handling very large datasets.}
\item \za{\textbf{DG3. Support Multiple AI Evaluators.} \texttt{EvalAssist} provides multiple AI evaluators. Users can select their preferred model as the evaluator, addressing issues like self-enhancement bias \cite{xu2024pride}, where models favor their own responses. EvalAssist is model-agnostic, enabling users to choose any model for evaluation, unlike other front-end tools that limit users to a single AI evaluator.}
\item \za{\textbf{DG4. Include Bias Indicators.} ML researchers using LLM-as-a-judge are aware of potential biases exhibited by AI evaluation and have expressed the desire to have transparent indicators on bias in LLM-as-a-judge tooling \cite{pan2024human}. \texttt{EvalAssist} is the only front-end LLM-as-a-judge tool that provides a positional bias indicator. While this tool specifically focuses on positional bias, it sets the stage for addressing other issues like self-enhancement \cite{xu2024pride} and verbosity bias \cite{saito2023verbosity,ZhengChiangSheng2024} in the future. Including bias indicators helps users recognize positional biases in LLM judgments. This version explicitly flags positional bias, marking the first instance of an evaluation tool highlighting bias for users. It also paves the way for future exploration of additional biases and their impact on user responses.}
\item \za{\textbf{DG5. Enable Flexible Evaluation Methods} \texttt{EvalAssist} is the first tool to allow users to define their criteria for the two most common LLM-as-a-judge approaches: direct assessment (score-based, single-grading) and pairwise comparison (relation-based) \cite{ZhengChiangSheng2024,shi2024judging}. With this tool, users can select the strategy that best suits their dataset or explore both approaches to determine which is more effective for their needs.}

%\item \za{\textbf{DG6. Establish Shared Vocabulary with AI Evaluator.}   Users evaluating AI outputs often face challenges in defining criteria for tasks involving multiple relevant artifacts \cite{pan2024human}. Elements to be evaluated can range from simple single-input, single-output evaluations (e.g., answering a question) to complex scenarios where outputs depend on a combination of elements, such as a dialog, a reference article, and/or a question-answer pair. \texttt{EvalAssist} introduces a codebook-driven approach that allows users to define multiple variables for task-specific artifacts, enabling explicit and consistent references in criteria definitions. Existing tools enable defining commands or prompts \cite{arawjo2024chainforge} or variables within instruction prompts \cite{kim2023evallm}, but they lack support for referencing these variables in criteria to establish a shared vocabulary with the AI evaluator.}

\end{description}

\section{EvalAssist: System Design} 
%\mike{I was initially confused by this first statement. It may be better to say that \texttt{EvalAssist}}.

\texttt{EvalAssist} abstracts the llm-as-a-judge evaluation process into a library of parameterize-able evaluators (the criterion being the parameter), allowing the user
to focus on criteria definition. This approach acknowledges that developers often use complex external workflows to adjust configurations and experiment with different models and prompts to generate responses \cite{desmond2024evalullm}.  \texttt{EvalAssist} consists of a web-based user experience, an API, and a Python toolkit. The user interface  provides users with a convenient way of iteratively testing and refining LLM-as-a-judge criteria, and supports both direct (rubric-based) and pairwise assessment paradigms (relation-based) (Figure \ref{fig:evaloffering}), the two most prevalent forms of LLM-as-a-judge evaluation available \cite{kim2023prometheus,ZhengChiangSheng2024}. \za{\texttt{EvalAssist} is designed to be model-agnostic, allowing any instruction-tuned model to be used as the evaluator. Our system leverages the ability of models to follow instructions and supports four judges: \texttt{mixtral-8x7b-instruct-v01}, \texttt{llama-3-8b-instruct}, \texttt{llama-3-70b-instruct}, and \texttt{prometheus-8x7b-v2}. These models were selected because they represent state-of-the-art open-source instruction-tuned language models available at the time of the study and demonstrated strong performance according to HELM benchmarks \cite{liang2022holistic, helm}. Additionally, \texttt{Prometheus} was chosen specifically for its notable performance as an evaluator model in prior studies \cite{kim2023prometheus}.}

% \texttt{EvalAssist} supports Direct Assessment and Pairwise assessment in \texttt{EvalAssist}because they are the most prevalent forms of LLM-as-a-judge evaluation paradigms currently available \cite{kim2023prometheus,ZhengChiangSheng2024}. 

\begin{figure*}
    \centering
 \includegraphics[width=1\linewidth]{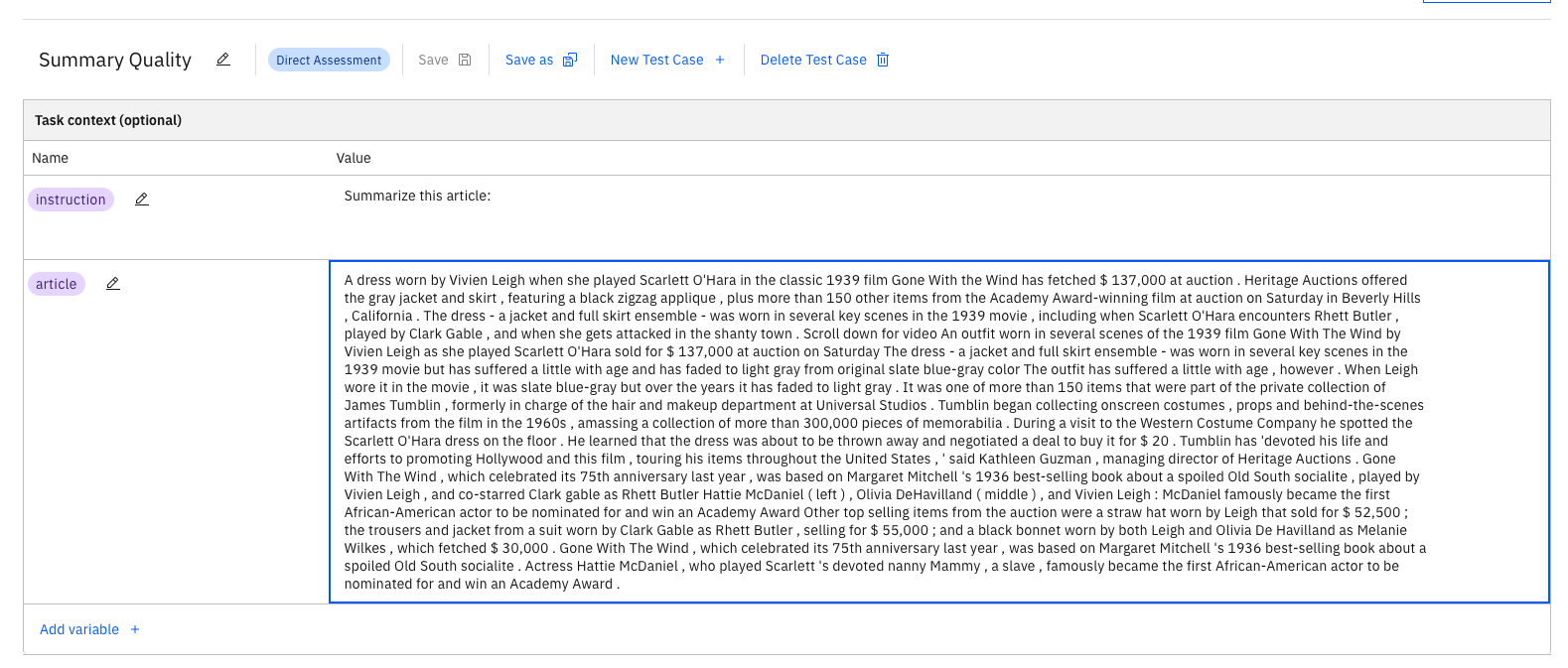}
    \caption{Task Context for the Summarization Task. The Task Context is consistent for both direct assessment and pairwise comparison strategies. Users have the option to break down the context into variables, such as the instruction and article, to simplify reference while developing evaluation criteria.}
    \label{fig:taskcontext}
\end{figure*}

Users can choose the evaluation method based on task complexity, receive AI judgment explanations, and view metrics like positional bias. Once users are satisfied with their criteria, they can use the Python toolkit to run bulk evaluations with larger data sets by exporting auto-generated JSON definitions of their criteria into predefined notebooks provided in the toolkit. We also allow users to save their test cases and provide a catalog of predefined criteria. A test case in the Example Catalog includes a criteria definition and the data being evaluated. On the landing page, users can choose between direct assessments and pairwise comparisons. \za{To be able to refer to artifacts and documents relevant to criteria generation (i.e., the output is relevant to the `dialog' or faithful to the `reference document'), users can define task-relevant input data through variables in the task-context (Figure \ref{fig:taskcontext}), such as, for example, the prompt, the article to summarize, or the source data for content-grounded Q\&A. This feature makes it easier to reference these elements when defining criteria in the evaluation form (Figure \ref{fig:sidebyside}).} \zauist{Outputs evaluated in EvalAssist may originate from prompts that are not visible to participants. Depending on task design, prompts and associated context can either be shown or withheld. This separation between prompt generation and evaluation supports the study of how criteria generalize across unseen prompts and reinforces the importance of distinguishing prompt-specific tuning from evaluation criteria.}

%Below are the core components and features of \texttt{EvalAssist}.
%The system is comprised of core modules: 
%\begin{itemize}
%\end{itemize}

\subsection{Direct Assessment}
\begin{figure*}[ht!]
    \centering
    \begin{subfigure}[b]{0.49\textwidth}
        \centering
        \includegraphics[width=\textwidth]{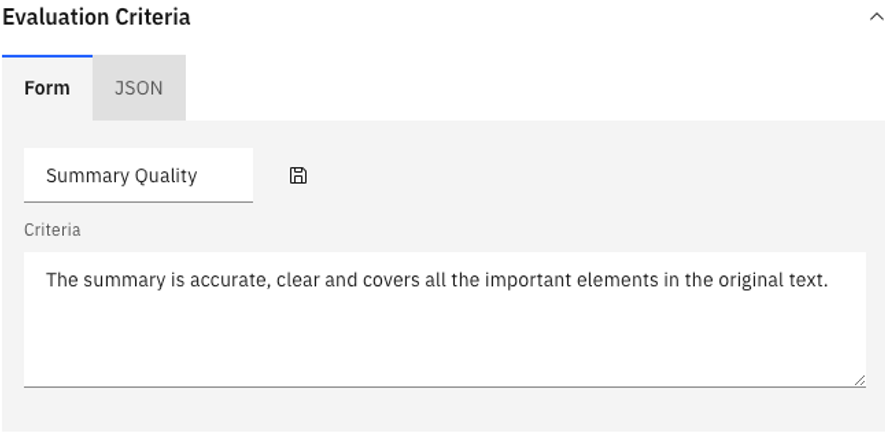}
        \caption{Pairwise approach: Features a concise one-sentence summary of each criterion.}
        \label{fig:figure1}
    \end{subfigure}
  \hspace{0.001\textwidth}
    \begin{subfigure}[b]{0.49\textwidth}
        \centering
       \includegraphics[width=\textwidth]{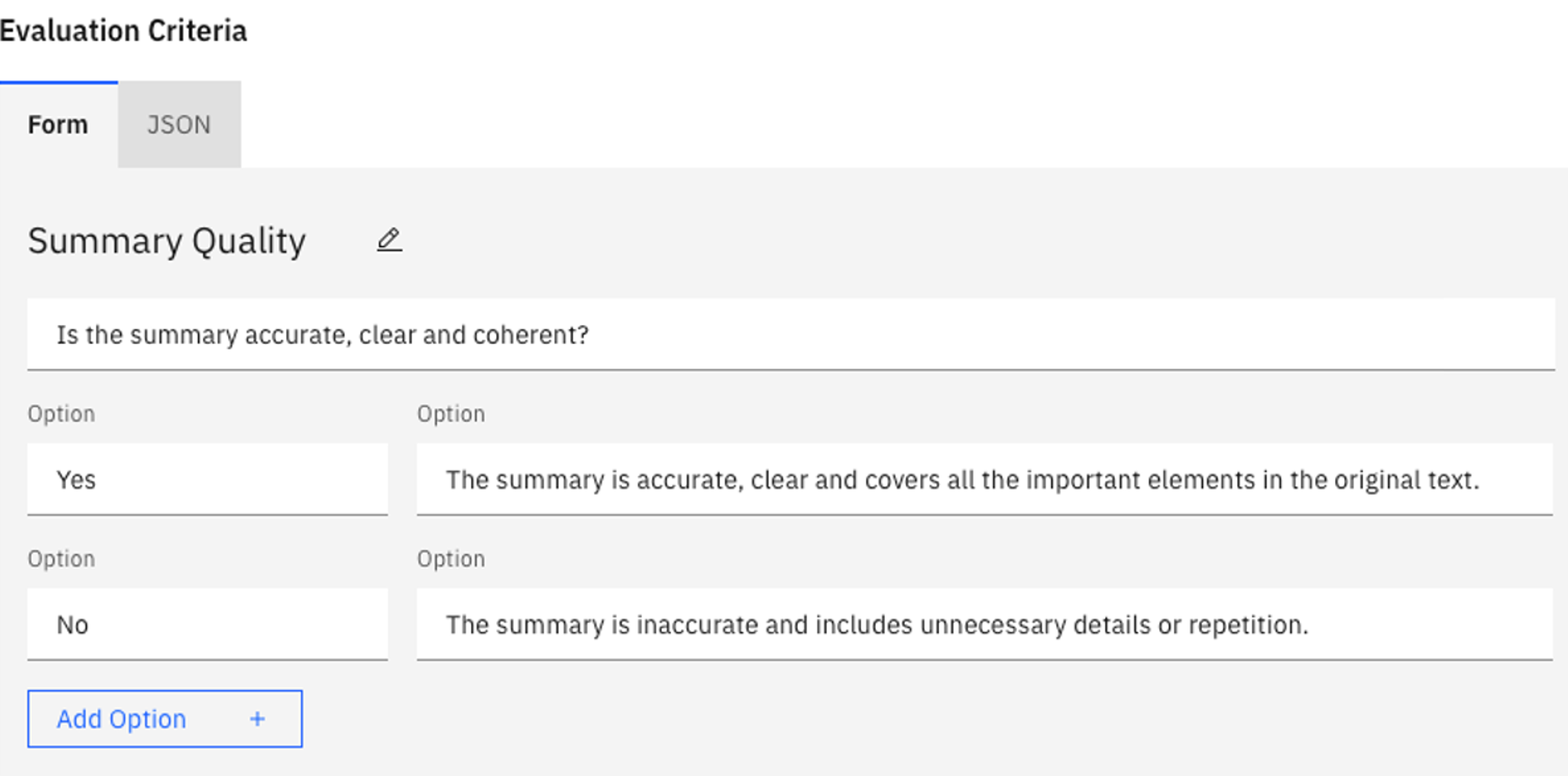}
        \caption{Direct assessment: contains a high-level question, scale items, and detailed optional definitions for each scale item.}
        \label{fig:figure2}
    \end{subfigure}
\caption{Evaluation criteria forms for pairwise assessment and direct assessment.}
    \label{fig:sidebyside}
\end{figure*}

 In this mode, users evaluate outputs based on a single criterion rubric they define. The evaluation criteria forms allow them to define their criteria with a title, criteria description, and an arbitrary number of free-form options the AI Evaluator will have to choose from during assessment (Figure \ref{fig:figure2}). As such, the 
 %Users can create multi-dimensional rubrics with specific criteria, which can range from objective (e.g., grammatical accuracy) to subjective (e.g., creativity). 
 system supports both binary and multi-level scale assessments. In the Evaluator section, users need to select an AI evaluator.  In the Test Data Section, users enter the outputs they want to evaluate (we call them the responses; note that it is possible to edit the variable name here too so the data being evaluated can be better referenced in the criteria definition) and optionally the result they would expect for each output. After running the evaluation, the system shows the actual results next to the expected results, including agreement, positional bias if present, certainty scores, and an explanation.

\begin{figure*}
    \centering
 \includegraphics[width=1\linewidth]{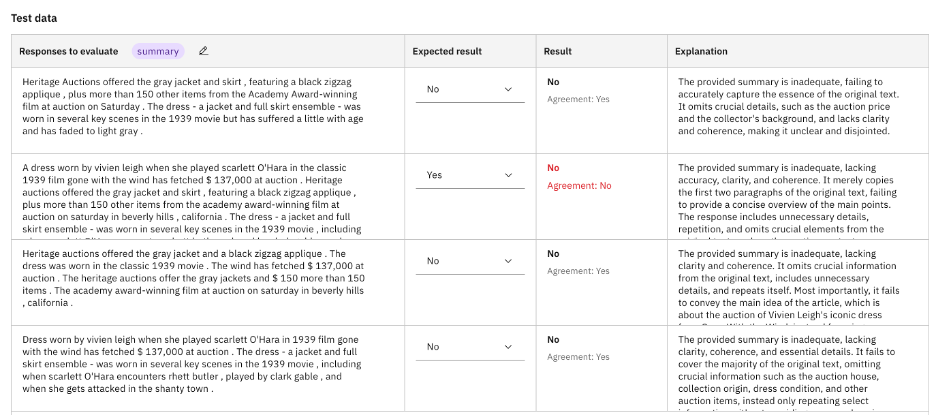}
    \caption{Results for direct assessment. Users can select their expected judgments for the output, which are auto-populated based on the criteria they define (i.e., the scale items created when setting the criteria). The results display the AI Evaluator's judgments, indicating whether there is agreement between the user and the AI, along with explanations for each result.}
    \label{fig:rubricresults}
\end{figure*}

\subsection{Pairwise Comparison}
\begin{figure*}[h!]
    \centering
    \begin{subfigure}[b]{0.68\textwidth}
        \centering
        \includegraphics[width=\textwidth]{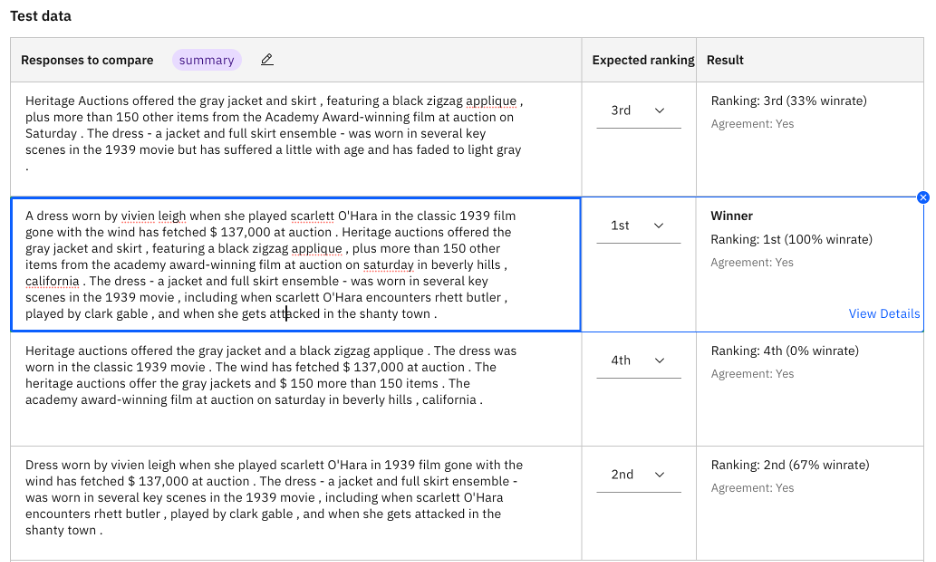}
        \caption{Ranking results generated from pairwise comparison assessment. Users can input their expected ranking to and assess their level of agreement with the AI evaluator.}
        \label{fig:ranking2}
    \end{subfigure}
  \hspace{0.01\textwidth}
    \begin{subfigure}[b]{0.28\textwidth}
        \centering
       \includegraphics[width=\textwidth]{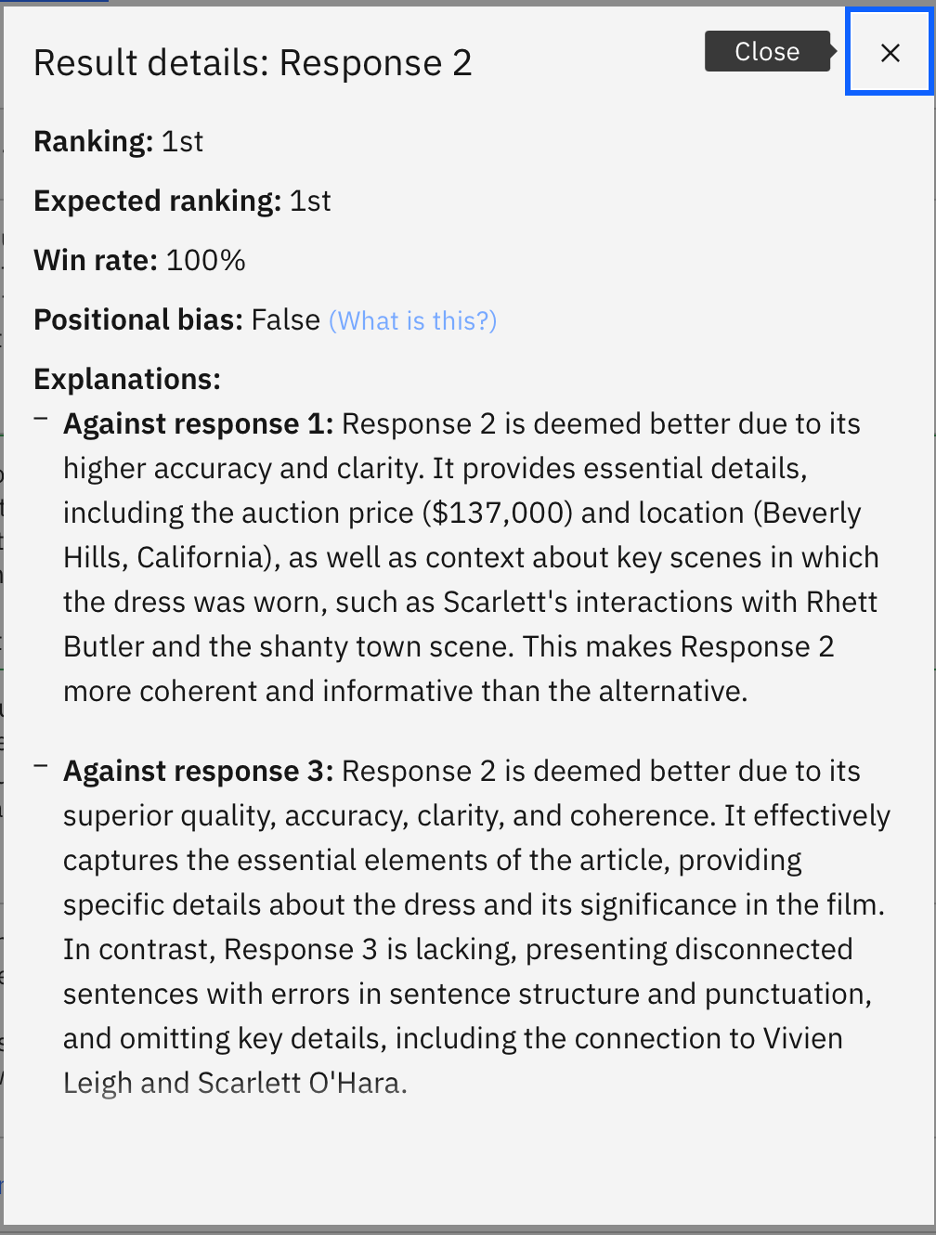}
        \caption{Explanations for each pairwise comparison in pairwise assessment.}
        \label{fig:figure2b}
    \end{subfigure}
    \caption{Results, explanations, and expected ranking generated through pairwise comparison. }
    \label{fig:sidebyside2}
\end{figure*}

In the pairwise comparison mode (Figure \ref{fig:sidebyside} (a)), \texttt{EvalAssist} compares multiple outputs (minimum two) pairwise against one-another selecting the one that better fits the criteria. The best output is determined by computing the win rate across all pairwise output comparisons. Similar to direct assessment, users can provide task-relevant input data through variables, define a criteria, and select an evaluator LLM. However, options don't need to be added to pairwise comparisons. After evaluation, we display the results next to the expected results (see Figure \ref{fig:sidebyside2}), including the winner, ranking, and agreement with expected ranking. \za{Explanations in pairwise comparison can be seen in Figure \ref{fig:figure2b} and are generated as a result of comparing pairs of responses to be evaluated. In total, $\binom{N}{2}$ comparisons are performed in a pairwise manner, where N is the total number of outputs being evaluated. Each pairwise comparison generates an explanation. The outputs are then ranked based on a win rate metric, similar to \cite{dubois2024alpacafarm}. The explanations presented in Figure \ref{fig:figure2b} correspond to the comparisons of each summary against the highlighted row in Figure \ref{fig:ranking2}. As a result,} users are able to click on each result to see detailed explanations including positional bias, win-rate, and explanations for the comparisons with the other outputs.

\subsection{\za{Positional Bias}}
\za{Positional bias occurs when a model consistently favors one option based solely on its position, rather than its content \cite{li2024split}. In evaluation tasks, such as pairwise comparisons, this bias arises if the model disproportionately selects the option in a particular position (e.g., always favoring the first option) regardless of the actual content. In direct assessments, where a response is evaluated based on specific criteria, the order of these options can be shuffled. To test for positional bias, the evaluation is conducted twice with the options presented in different orders. If the outcomes differ between the two evaluations, positional bias is present, as it indicates that the position of the options influenced the evaluation. Conversely, if the outcomes are the same, it suggests that the model's assessment was not influenced by the position. In \texttt{EvalAssist}, we include a positional bias indicator for both direct and pairwise assessments. This indicator is displayed for each row in the results and is flagged in red text to highlight inconsistent judgments by the AI evaluators.} 

%In essence, positional bias means that the location of the item being evaluated impacts the results.

\subsection{Evaluation}
\begin{figure*}
    \centering
 \includegraphics[width=.8\linewidth]{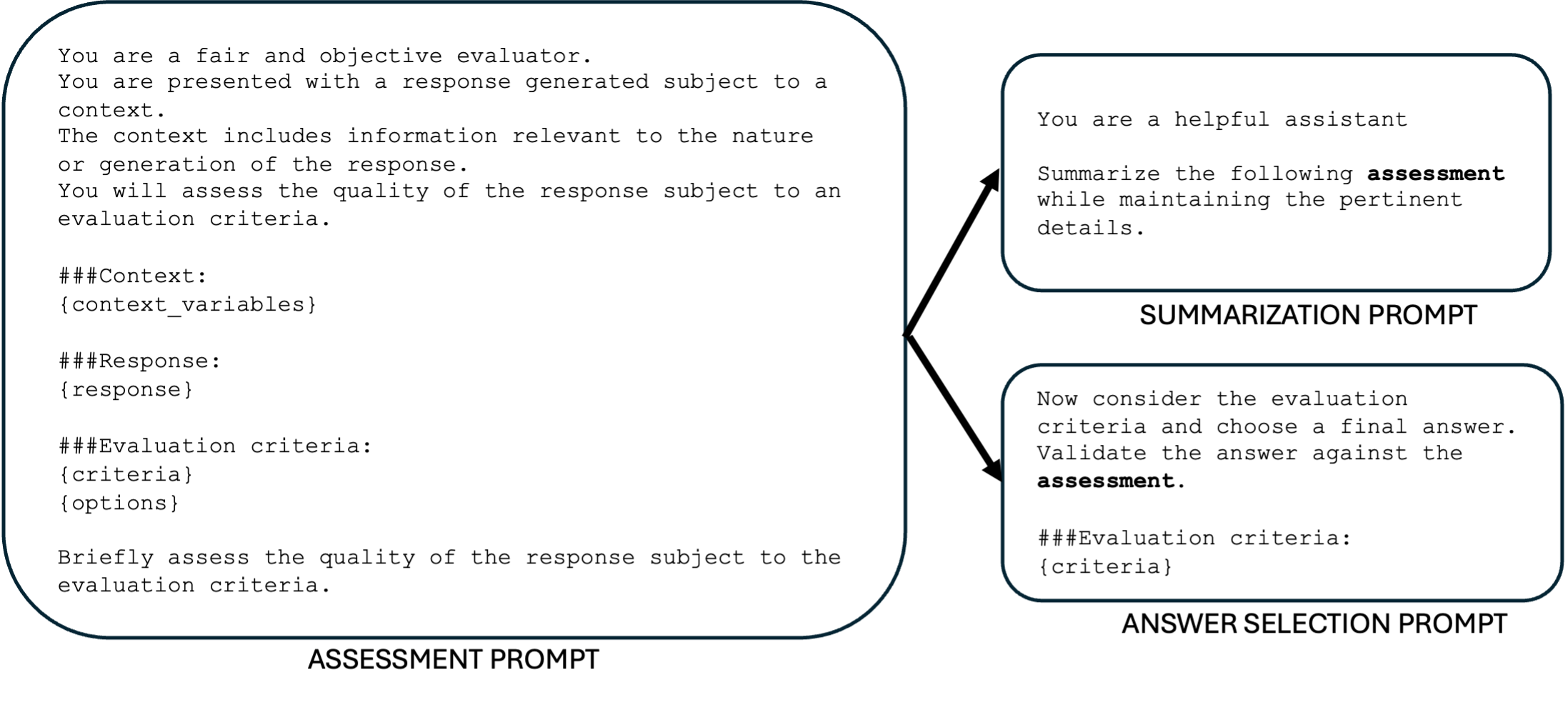}
    \caption{\za{The assessment prompt, summarization prompt and answer selection prompt for the \texttt{llama-3-8b-instruct} and \texttt{llama-3-70b-instruct} AI evaluators in the direct assessment context. The assessment prompt is used first, generating an evaluation of the response based on context and evaluation criteria. This assessment is then passed to a summarization prompt, which condenses the findings, and an answer selection prompt, which validates the response against the evaluation criteria.}}
    \label{fig:prompts}
\end{figure*}

When users select the "Evaluate" button, their input is sent to the chosen evaluator. Each evaluator is designed to perform either direct assessment or pairwise evaluation. The main external difference between these two lies in how the input criteria is structured. Internally, evaluators operate as a dialog with a target \za{AI evaluator} using a set of custom prompts specific to that \za{AI Evaluator. The process consists of an assessment prompt, a summarization prompt, and an answer selection prompt for all AI evaluators. However, \texttt{prometheus-8x7b-v2} uses only a single prompt, as it has already been trained to provide high-quality explanations \cite{kim2024prometheus}. Figure \ref{fig:prompts} demonstrates these prompts for the \texttt{llama-3-8b-instruct} and \texttt{llama-3-70b-instruct} AI evaluators. We include the remaining prompts for each AI evaluator in the Appendix. First,} the LLM is prompted to review the evaluation task, considering the context, criteria, and subject of evaluation. The LLM generates an open-ended assessment that explains its decision-making process. This step is inspired by Chain-of-Thought (CoT) prompting \cite{wei2022chain}, encouraging the LLM to base its final judgment on its initial reasoning.  This generated assessment is then added to the dialog history. Afterward, the LLM is asked to make a final judgment. Instead of having the LLM directly generate a final decision, we provide a set of options as potential completions. The system \za{extracts} the log probability of each option's tokens using a forward pass and selects the option with the highest probability. This comparison is done by evaluating the first token of each completion, followed by the second, and so on. For direct assessment, these completions are the user-provided option strings, while in pairwise evaluation, the completions are the two responses being compared (Response 1 and Response 2). Although using log probabilities for determining the LLM's judgment is less efficient than direct generation, it is significantly more reliable \za{because it eliminates the possibility of a hallucination which would lead to a failed evaluation}. \zauist{We use log probabilities over pre-generated completions as a proxy for model confidence, allowing for token-level diagnostics and detection of overconfident errors such as confidently hallucinated or off-topic content. This approach requires access to model logits, though \texttt{EvalAssist} can also operate in generation-based workflows when logits are unavailable. Prior work has shown that token-level confidence is a useful signal for identifying hallucinations \cite{10.1007/978-3-031-86623-4_13, kryscinski-etal-2020-evaluating}, and we draw on these insights to enhance the transparency of AI judgments.} The \za{summarization prompt results in a summarization of the initial assessment which serves as the explanation presented to the user.}  Positional bias is checked by shuffling the order of the options presented to the LLM and verifying the consistency of its final decision. A visual representation of the algorithm is shown in Figure \ref{fig:evalassist} in the Appendix. \zauist{For both pairwise and direct assessment tasks, \texttt{EvalAssist} includes positional bias analysis as an indicator of evaluation robustness. While positional bias is more commonly associated with pairwise tasks, it can also affect direct assessments, where the ordering of rating options (``Accurate''-``Inaccurate'' scale) may influence outcomes. \texttt{EvalAssist} supports inversion of these options to detect such order effects, and we expose this information to users as part of the judgment result interface.}

\section{Experimental Design}
\za{To better understand how people use EvalAssist to define criteria and evaluate output, we conducted a study exploring several key questions.} Specifically, \za{we sought to understand how participants approach the development and refinement of evaluation criteria when using LLMs as judges, focusing on the strategies, characteristics, and evaluative dimensions that emerge in their criteria creation process} (RQ1). The study also focused on the impact of task-related factors and judge strategies on how practitioners refine criteria (RQ2). Furthermore, the research investigated how task and judge strategy affect user perceptions of the AI judge, including trust, perception of positional bias, perception of explanations, and cognitive load (RQ3). Lastly, the study aimed to identify which judgment strategy users preferred (RQ4). 

We used a within-subjects study design, involving 15 participants recruited internally from our organization. Participants were recruited internally from individuals who had previously used \texttt{EvalAssist} and through announcements posted on the organization’s Slack channels. These messages invited employees to participate in a study, with eligibility focused prior experience with model evaluation. \za{The study followed company policies on user data. Participation was voluntary, with the option to withdraw anytime. Data was de-identified, stored securely, and analyzed anonymously. Participants consented to the use of de-identified data for current and future research, ensuring confidentiality.} Participants were provided with detailed information about the study's purpose, procedures, and their rights as participants.  We collected demographic information, including participants' roles at the company, education levels, and prior experience with model evaluation. Participants then completed a practice task to familiarize themselves with the experimental interface and procedures.

The experimental tasks involved creating criteria and evaluating output based on their criteria using either the pairwise approach or the direct assessment approach for six task contexts based on 3 tasks (Q\&A, Email Generation, Summarization) each seen twice, once under the direct assessment condition and once under the pairwise condition. \za{For each task, participants were instructed to stop the evaluation process when they felt satisfied with the criteria they had created. This approach provided them with autonomy and flexibility to determine when their criteria met their own standards or expectations.} Each task is described in Table \ref{tab:tasks}.

%Direct Assessment and Pairwise assessment are the most prevalent forms of evaluations \cite{kim2023prometheus,ZhengChiangSheng2024}. 
Direct assessment involved designing a single criterion rubric with options to assess whether output complies with the criteria (as shown in Figure \ref{fig:rubricresults}), whereas pairwise comparison (see Figure \ref{fig:sidebyside}) involves defining the criteria and comparing pairs) of generated outputs to see which better matches  criteria. While both approaches come with strengths and weaknesses, we set out to examine how they influence the evaluation of criteria and users' interactions differently. 

Each task was followed by a short survey assessing participants' trust in the AI, satisfaction with the criteria, cognitive load, perception of positional bias, and the explanations provided by the AI. Throughout the study, data were logged on the number of evaluations run, final human agreement with the AI Evaluator, and the time taken to run each evaluation. The following questions were rated on a 5-point Likert scale, with 1 indicating ``strongly disagree'' and 5 indicating ``strongly agree'':

\begin{itemize}
\item Trust in AI Evaluator:
\begin{itemize}
\item I trusted the AI evaluator to judge the responses. (adapted from \cite{poursabzi2021manipulating})
\item How confident were you in the model’s judgments/evaluations? (adapted from \cite{buccinca2020proxy})
\end{itemize}
\item Perception of Positional Bias: The positional bias was helpful in completing this task. 
\item Perception of Explanations: The explanations were helpful in completing this task. 
\item Mental Load (adapted from \cite{hart2006nasa}):
\begin{itemize}
    \item The task was mentally demanding. 
    \item I was successful in accomplishing what I was asked to do. 
    \item I had to work \zauist{hard} to accomplish my level of performance 
\end{itemize}
\end{itemize}

To mitigate order effects, we used partial counterbalancing. Participants were randomly assigned to different task orders to ensure that the sequence of tasks did not systematically bias the results. Participants were asked to reflect on the different tasks they had completed and respond to questions about their preferences between direct assessment and pairwise assessment. They also described how they interacted with two types of tasks: objective versus subjective. After completing the six tasks, participants \za{had a reflection phase in which they} were asked to select their preferred judgment strategy. \za{During the reflection phase, participants reviewed the completed tasks, focusing on differences in judgment strategies and task outcomes. They were asked to identify their preferred strategy and explain the reasoning behind their choice.}

\subsection{Tasks and Evaluation Criteria}
Tasks and evaluation criteria were chosen to capture various levels of granularity and criteria specificity (i.e., single dimension of `preference' vs. document groundedness) and to reflect a diversity of tasks and respective models. \za{Our goal was to design criteria that reflect real-world tasks practitioners encounter, which naturally encompass a diversity of task types. We intentionally avoided defining the criteria as ``ambiguous vs. specific'' or ``subjective vs. objective'' because this framing did not align with the variety of tasks included. Instead, we focused on understanding the differences in how participants approached these tasks, specifically, alignment, user preferences, and the types of evaluations or behaviors they exhibited.} Responses were drawn from a variety of sources (models, datasets), and were reviewed by the coauthors to ensure sufficient variability. Below we list the task descriptions. Task examples can be seen in Table \ref{tab:tasks}. 

\begin{itemize}
\item \textbf{Article Summarization: } While the summaries can be judged on cohesiveness, consistency, fluency, relevance \cite{Bavaresco2024JUDGE_BENCH} we asked participants to define criteria based on the single dimension of ``preference'', as seen in work by \cite{chiang2023can,li2023alpacaeval,liu2023g}. For the summarization task, we presented the original reference document and corresponding generated outputs to users to be judged from \cite{fabbri2021summeval}. 
\item \textbf{Email Generation}: The email generation task was leveraged to judge  inclusivity. We generated an email about an office Christmas party using various models (Gemini 1.5-Pro \cite{reid2024gemini}, claude-3-5-sonnet-20240620 \cite{anthropic2024claude35}, gpt-3.5-turbo-0125 \cite{floridi2020gpt}, mixtral-8x22b-instruct-v0.1 \cite{jiang2023mistral}), resulting in emails with different levels of inclusivity. We asked participants to create criteria to evaluate the generated output based on the inclusiveness of the output. 
%\item \textbf{Audio Transcription}: We asked users to judge transcribed audio based on accuracy. The task was selected from \cite{Bavaresco2024JUDGE_BENCH} and the instructions included \textit{Please reproduce this report verbatim, but do correct where items have been spelled out verbally (e.g. Cal gary C A L G A R Y -> Calgary)}. Because the source of the dataset \cite{Bavaresco2024JUDGE_BENCH} only provided two model outputs, we regenerated other outputs for our tasks with mixtral-8x22b-instruct-v0.1, gpt-4o-2024-05-13, claude-3-5-sonnet-20240620, dbrx-instruct. 
\item \textbf{Q\&A multi-turn:} The Q\&A task involved a context document, a multi-turn conversation, and a final response to the last question in the conversation. The output was generated using retrieval-augmented generation (RAG) \cite{lewis2020retrieval}. RAG evaluations focused on three aspects: answer relevance (is the answer relevant to the query?), context relevance (is the retrieved context relevant to the query?), and groundedness (is the response supported by the context?) \cite{yu2024evaluation}. The Q\&A data was selected from existing HR support documents at our company, with AI-generated responses created by multiple models, including \cite{https://doi.org/10.48550/arxiv.2210.11416}. One of the question-and-answer sets involved a reference document and a customer inquiry about upgrading their business travel to business class with their corporate credit card. Participants were asked to create a criterion to evaluate the faithfulness of each response to the reference document. A response is considered faithful to a document when it only reflects the information expressed in the document \cite{madsen2024self}. 
\end{itemize}

\begin{table*}[h]
\footnotesize
\centering
\begin{tabular}{@{}p{3.2cm}p{12.8cm}@{}}
\toprule
\textbf{Task Description} & \textbf{Example Output} \\
\midrule

\textbf{Article Summarization} &
Dress worn by Vivien Leigh when she played Scarlett O’Hara in the 1939 film *Gone with the Wind* has fetched \$137,000 at auction. \newline
The dress—a jacket and full skirt ensemble—was worn in several key scenes in the movie, including when Scarlett encounters Rhett Butler, played by Clark Gable, and when she is attacked in the shanty town. \\

\addlinespace[0.5em]

\textbf{Q\&A in HR Chatbot} &
You can upgrade to business class by paying the business class portion of the airline ticket with a personal credit card to avoid the perception of a policy bypass. \\

\addlinespace[0.5em]

\textbf{Email Generation} &
Subject: Join Us for a Multicultural Holiday Celebration! \newline
Dear Team, \newline
As the holiday season approaches, we are thrilled to announce our annual company holiday party! This year, we are embracing the diverse cultures within our team and celebrating a variety of holiday traditions. \newline\newline
Details are as follows: \newline
Date: December 20th \newline
Time: 6:00 PM – 10:00 PM \newline
Location: Main Conference Hall \newline\newline
Our celebration will feature a variety of activities that honor different cultural traditions, including:\newline
– Decorations from Around the World: Help us decorate the office with symbols and ornaments from various holidays, such as Christmas trees, Hanukkah menorahs, Kwanzaa kinaras, Diwali lamps, and more. \newline
– Inclusive Gift Exchange: Participate in a gift exchange that embraces the spirit of giving across cultures. Bring a wrapped gift (under \$20) or a small item that represents your cultural heritage. \newline
– Global Holiday Music: Enjoy a diverse selection of holiday music, including Christmas carols, Hanukkah songs, Kwanzaa tunes, and traditional music from different cultures. \newline
– International Feast: Indulge in a delicious array of holiday dishes from around the world. We will have a variety of foods to cater to different tastes and dietary preferences. \newline\newline
Feel free to wear traditional attire or festive clothing that reflects your cultural heritage. Whether it’s a Christmas sweater, traditional clothing for Hanukkah, Kwanzaa, Diwali, or any other holiday, we welcome it all! \newline\newline
We would love to hear your ideas and suggestions to make this event even more inclusive and enjoyable. If you have any special traditions or activities you’d like to share, please let us know. \newline\newline
Let’s come together to celebrate the season and the wonderful diversity within our team. \newline\newline
Happy Holidays! \newline
Best regards, \newline
Company CEO \\

\bottomrule
\end{tabular}
\caption{Example outputs generated for three distinct tasks in the evaluation study: Article Summarization, Q\&A in an HR Chatbot, and Email Generation.}
\label{tab:tasks}
\end{table*}

\section{Results}

\subsection{Participant Demographics}
We recruited 15 participants at our company who had prior experience with model evaluation. The job titles are listed in Table \ref{tab:participants}. Their prior experience included subjective and manual evaluation, use case specific evaluation, evaluation framework and metrics. 
%\zauist{One limitation of this study is that participants were recruited from a single organization. While they had relevant domain experience, this sample may limit the generalizability of our findings. Future studies should include a broader and more diverse participant pool from external sources to validate and extend these insights.}

\begin{table}[h!]
\small
\centering
\begin{tabular}{c|l}
\hline
\textbf{ID} & \textbf{Job Role}  \\ \hline
P1 & AI Engineer  \\ \hline
P2 & Research Scientist Intern \\ \hline
P3 & Software Developer \\ \hline
P4 & Data Scientist \\ \hline
P5 & Software Engineer\\ \hline
P6 & Senior Research Scientist \\ \hline
P7 & Advisory AI Engineer \\ \hline
P8 & Senior AI Technical Architect \\ \hline
P9 & Distinguished Engineer and Master Inventor \\ \hline
P10 & Research Scientist \\ \hline
P11 & Research Software Engineer \\ \hline
P12 & Senior Technical Staff Member \\ \hline
P13 & Platform Engineer \\ \hline
P14 & Research Scientist Intern\\ \hline
P15 & Technology Engineer \\ \hline
\end{tabular}
\caption{Participant IDs and job roles}
\label{tab:participants}
\end{table}

\begin{quote}
\textit{As a technology engineer, I interact with various LLM to perform many generative-ai tasks. However, besides manual inspection of LLM outputs, I have not found a way to consistently evaluate the performance of a LLM and see if each prompt is performing better or worse. }
\end{quote}

\begin{quote}
\textit{I've had to look at a number of tooling before in relation to checking the validity of models. (Cross validation, Lime, SHAP, Rouge, Bleu, Flesch-Kincaid, Flesch.  I've played with others but not to a level I can say I would know enough about them (eg. SQuAD, CIDEr, SPICE, WIT, ELI5) }
\end{quote}

\subsection{RQ1: How do participants approach the development and refinement of evaluation criteria when using EvalAssist, and what strategies, characteristics, and evaluative dimensions emerge in their criteria creation process?}
\za{In our first research question, we are interested in exploring the diversity of approaches leveraged by participants throughout the criteria development process. Because iteration is a key affordance of \texttt{EvalAssist}, we examined the strategies participants employed within tasks to refine their criteria (Table \ref{tab:changesmade}). Beyond the changes made during each evaluation within a task, we also investigated the characteristics of the criteria when participants reported satisfaction with their criteria at the completion of an evaluation task (Table \ref{tab:criteriacode}). Lastly, we explored how participants prioritized certain evaluative dimensions in their criteria (e.g., faithfulness versus conciseness) and how they translated task instructions into actionable criteria for the AI evaluator (Table \ref{tab:interpretations}). To generate these insights, we employed inductive coding \cite{thomas2006general}. Two coauthors independently extracted themes from the data, focusing on the characteristics of the criteria generated for each task. They then met to compare and discuss their initial codes to identify areas of agreement and disagreement. Discrepancies were resolved through discussion, and additional rounds of refinement were conducted to ensure that the codes accurately captured the nuances of the data. This process culminated in the development of a consensus codebook, which was iteratively refined and systematically applied to the entire dataset to ensure reliability and consistency.}

%\za{In our first research question we are interested in observing the diversity in approaches leveraged by our participants when throughout the criteria process. Because iteration is an important affordance of \texttt{EvalAssist}, we were interested in strategies users used within tasks to refine their criteria (Table \ref{tab:changesmade}). Beyond the changes made across each evaluation within a task, we were interested in the characteristics of the criteria when users were satisfied with their criteria upon the completion of an evaluation task (Table \ref{tab:criteriacode}. Lastly, we were interested in the prioritization on certain evaluative dimensions of the criteria users created and how they translated task instructions into criteria for the AI evaluator (Table \ref{tab:interpretations}). All of these codes were generated using inductive coding \cite{thomas2006general}. Two of the coauthors independently extracted themes from the data, focusing on characteristics of the criteria generated for each task. The authors then met to compare and discuss their initial codes to identify areas of agreement and disagreement. Discrepancies were resolved through discussion, and additional rounds of refinement were conducted to ensure the codes accurately captured the nuances of the data. This process led to the development of a consensus codebook, which was iteratively refined and applied to the entire dataset to ensure reliability and consistency.}

\subsubsection{Strategies for Refining Criteria}
\begin{figure*}
    \centering
 \includegraphics[width=.6\linewidth]{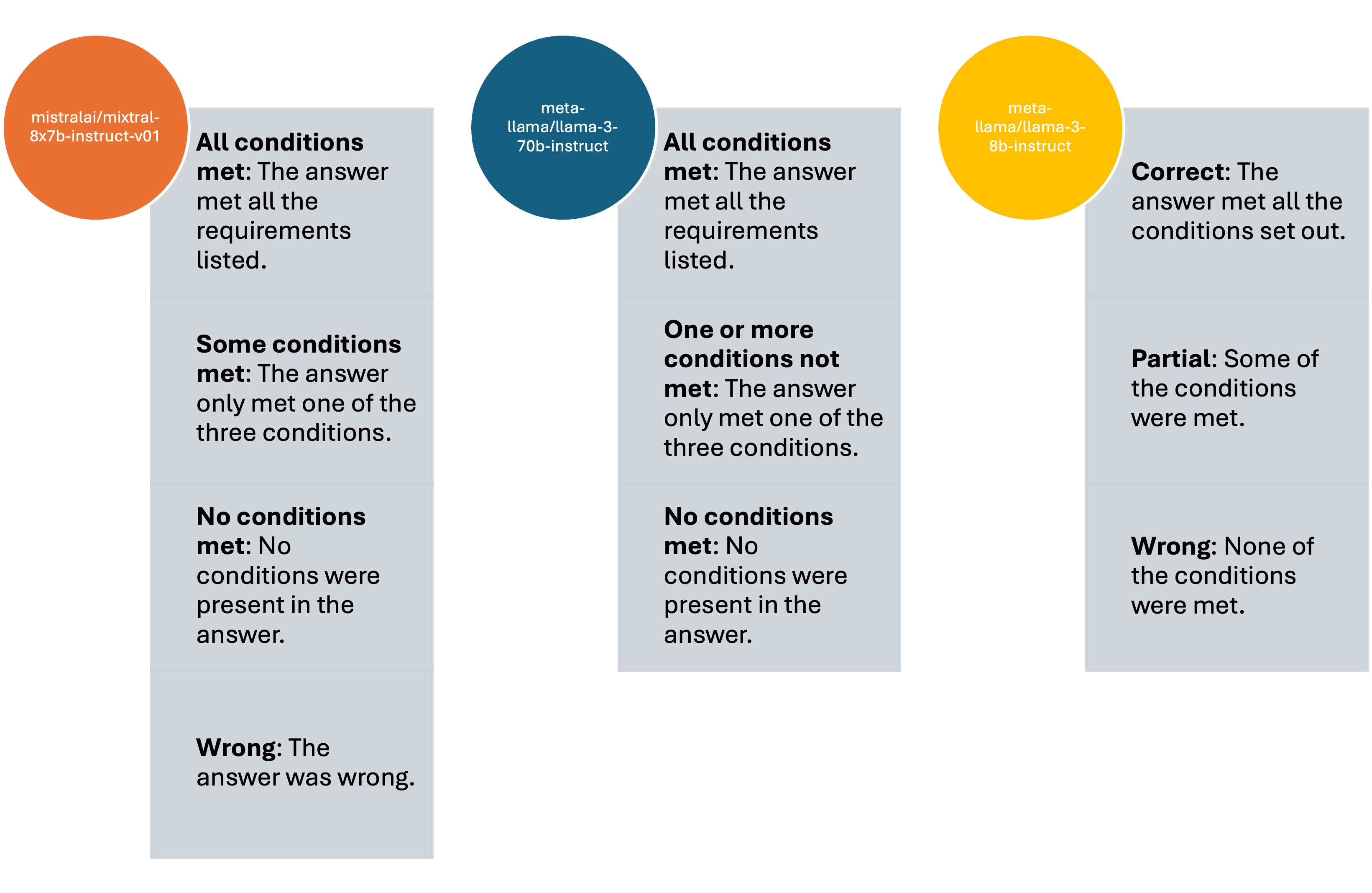}
    \caption{Example of changes made to criteria and AI evaluator for the by Participant \#1 for the Q\&A HR Task. The participant changed the AI evaluator model across the evaluations and changed the number of items in the scale as well as their corresponding definitions. The criteria definition remained the same throughout the evaluations: \textit{Did the answer give the following? (1) Factually correct answer based on the document. (2) Included related information directly to the question. (3) Answer is clear and concise.}}
    \label{fig:ev}
\end{figure*}

\begin{figure*}
    \centering
    \includegraphics[width=1\linewidth]{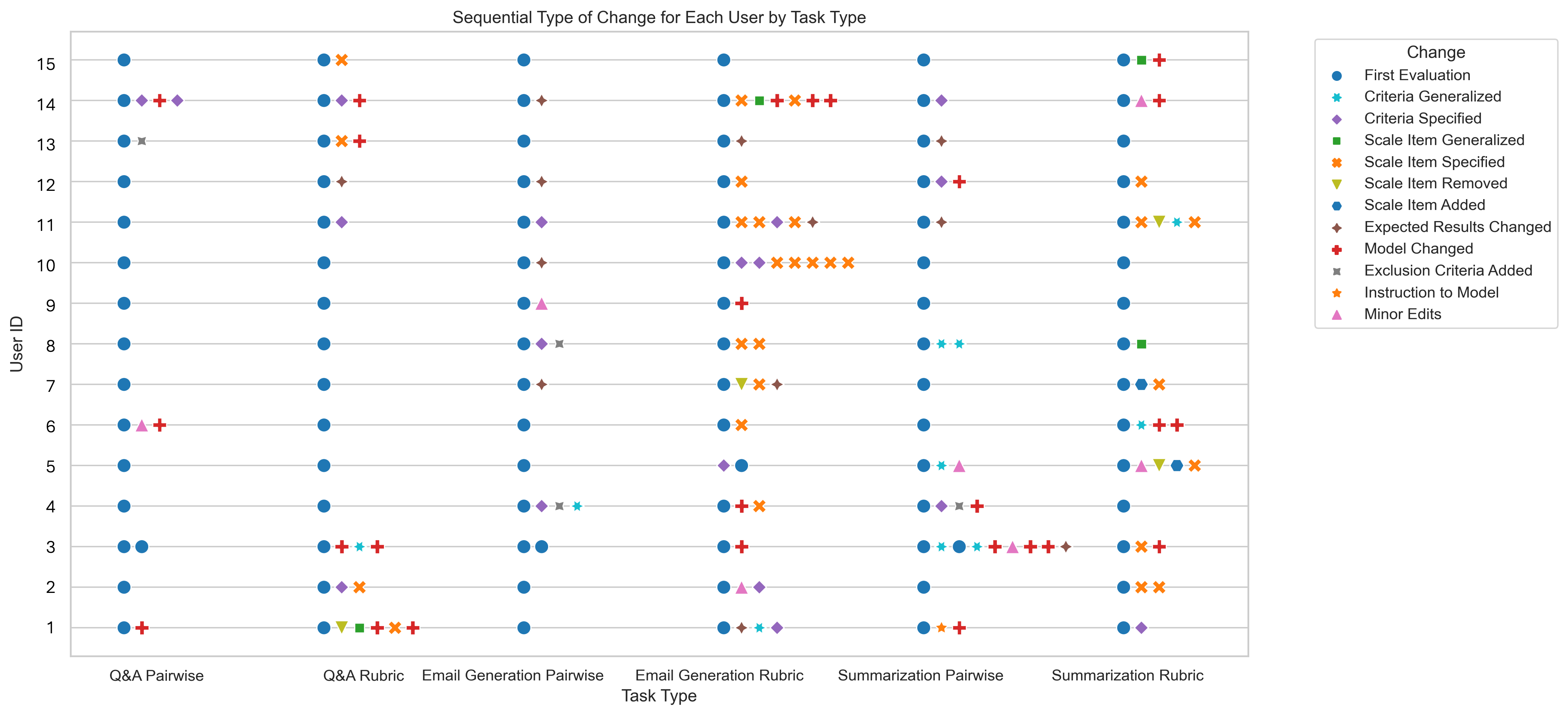}
    \caption{Sequential changes in evaluation criteria by task type and user. Symbols represent different types of modifications, including changes to criteria, scale items, models, and exclusions, with variation across Q\&A, Email Generation, and Summarization tasks.}
    \label{fig:sequence}
    \end{figure*}

\za{Table {\ref{tab:changesmade}} summarizes strategies used to refine criteria based on the coded meta-data from all logged evaluations. This data included the criteria and AI evaluators applied across various tasks.  In the direct assessment context, the most common refinement strategy was to make definitions more specific (37\%), followed by changes to the model (18\%). For example, in the Email Inclusivity task, a participant refined the criterion "The generated email should be inclusive" by specifying that it should use terms like "holiday" and "festive" instead of potentially non-inclusive language. This added precision offered clearer guidance for evaluation. Later, in Section {\ref{sec:specific}}, we explore these trends in detail, with a particular focus on how specificity influences criteria creation. Figure {\ref{fig:sequence}} depicts the sequential progression of changes.  Figure {\ref{fig:ev}} provides an example of how criteria evolved over time. }

In the direct assessment context, participants also made refinements by simplifying or elaborating on scale items. For instance, in the Summarization task, a multi-level scale ("Absolutely," "Somewhat," "No") was simplified to a binary scale ("Yes" or "No") for greater clarity. Conversely, some scales were expanded to include additional levels such as "Excellent," "Good," "Average," and "Poor." Additionally, participants clarified existing scale items by adding definitions. For example, in the Email Inclusivity task, they explicitly defined terms like "Inclusive" and "Not Inclusive." In contrast, some users generalized scale items, broadening them to account for a wider range of cultural practices.

One common modification was making criteria more specific. In contrast, some participants generalized their criteria to broaden their scope. In the same task, a criterion initially focused on specific groups, such as cultures and backgrounds, was simplified to a more general statement: "The email is inclusive to all cultures and backgrounds."
Minor edits were also common, such as rephrasing criteria for clarity. For example, the question "What is the best summary?" was revised to include clearer instructions: "Which of the following summaries best describe the article? Review each independently of the others." Some participants added exclusion criteria to prevent irrelevant elements from being considered. One participant, for instance, added a criterion stating that "The summary should be accurate and concise. Has no fake data generated outside of reference," clearly outlining what should be excluded. These refinements illustrate the range of strategies participants used to tailor their criteria, from increasing specificity to allowing greater flexibility. Figure \ref{fig:sequence} shows the sequential changes in evaluation criteria by task type and user.

\subsubsection{Characteristics of Finalized Criteria}
The \za{final} criteria that users created varied widely across tasks, not only in terms of specificity, but also in the presence of instructions or rules, the number of items in the scale (for the direct assessment condition), the inclusion of exclusion criteria, and the use of examples. These variations reflected what aspects users prioritized when defining their criteria (see Table \ref{tab:criteriacode}). We observed a range of specificity, with some users providing highly detailed and task-specific criteria, while others offered more general guidelines. In some cases, users added rules or additional prompting to guide the model. For example, some criteria included explicit if-then rules to distinguish between acceptable and unacceptable options. \za{This may be explained by users' need for transparency, specifically, their inability to see exactly how the AI evaluator is being prompted, leading them to add their own additional prompts. For example, the inclusion of, ``You must review each summary independently of other summaries when making your judgment.'' in the criteria suggests that the user believed the summaries were not being evaluated independently. We discuss providing transparency to users in Section \ref{sec:transparency} based on these findings.}

Additionally, some users incorporated examples within their criteria to illustrate preferred outcomes, effectively providing the model with concrete cases to guide its responses. The lack of transparency here fails to clearly communicate to the user that the model may not be incorporating such examples during the evaluation phase. This approach and others highlights the diverse ways in which participants interpreted and operationalized the constructs in question. 

\subsubsection{Prioritization of Evaluative Dimensions}
Within each task, participants \za{generated diverse criteria for each task reflecting varied prioritization of evaluative dimensions (see Table \ref{tab:interpretations}). Our tasks were purposefully diverse so that we can explore the varied prioritization across tasks. } 

In the Email Inclusivity task, participants prioritized cultural inclusivity, neutrality, and fairness. Some emphasized mentioning all cultures to avoid bias, ensuring holidays like Christmas, Hanukkah, Kwanzaa, and Diwali were included. \za{This approach reflected a belief in the importance of visibility for diverse groups.} Others focused on maintaining a neutral tone by avoiding specific mentions of culture, gender, race, or ethnicity. \za{Their goal was to maintain a universal tone that could resonate with a broader audience without alienation.} Additionally, some prioritized fairness, ensuring the email did not favor one culture over another. One participant intepreted inclusivity to mean financial inclusivity and designed the criteria around adherence to budget constraints, such as keeping gift exchange mentions within a \$40 limit. \za{By designing criteria that specified gift exchange limits (e.g., within a \$40 budget), this participant highlighted an often-overlooked dimension of inclusivity: financial fairness.}

In the Summarization task, participants varied in their focus on factual accuracy, inclusion of key points, and conciseness. Some prioritized ensuring the summary accurately reflected the article, while others focused on covering all key points comprehensively. Conciseness was also important, with participants aiming for a summary around 20\% of the original length. \za{Beyond these priorities, some participants leaned on grammatical correctness and the inclusion of specific details, such as dates or prices, to enhance clarity and utility.} Some participants relied on the AI to decide which is the best summary by keeping their criteria general and asking for the the ``best" summary. \za{This strategy reflected trust in the system's evaluative judgment.}

For the Q\&A Faithfulness task, participants focused on maintaining faithfulness to the source, adhering to policy, avoiding hallucinations, and ensuring correctness. They emphasized that responses should be strictly grounded in the reference document and follow company policy, with no deviation or fabrication.

\za{These varied approaches across tasks demonstrate the diversity in participant priorities and interpretations. Each decision reflected a combination of task-specific goals, individual values, and context-driven judgments, revealing the complexity of aligning evaluation criteria with multifaceted objectives. This richness in participant strategies highlights the importance of flexibility and transparency in criteria development to accommodate diverse perspectives.}

\begin{table*}[h!]
\small
\centering
\begin{tabular}{p{3cm}p{6cm}p{6cm}}
\toprule
\textbf{Type of Change} & \textbf{Original Criteria}&\textbf{Changed Criteria} \\ \midrule
Criteria Specified & \textit{The generated email should be inclusive. It should mention the different cultures and holidays (such as Christmas, Hanukkah, Kwanzaa, Diwali, and others) . It should not be exclusive to one culture. } & \textit{The generated email should be inclusive. It should mention the different cultures and holidays (such as Christmas, Hanukkah, Kwanzaa, Diwali, and others) . It should not be exclusive to one culture. The email should use inclusive terms such as holiday and festive as opposed to terms exclusive to one culture. }\\
\hline
Criteria Generalized & \textit{The email is inclusive to all cultures, backgrounds, genders, etc.} & \textit{The email is inclusive to all cultures, and backgrounds.}\\
\hline

Scale Item Removed& \textit{Does the summary contain the main topic of the article?} &\textit{Does the summary contain the main topic of the article?}\\
&Scale: Absolutely, Somewhat, No &Scale: Yes, No\\
\hline
Scale Item Added &\textit{Does the response capture the summary in the best possible way?}&\textit{Does the response capture the summary in the best possible way?}\\
&Scale: Yes, No & Scale: Excellent, Good, Average, Poor\\
\hline
Scale Item Specified &\textit{Read the following email and determine if the email is inclusive or not inclusive of cultural differences. }&  \textit{Read the following email and determine if the email is inclusive or not inclusive of cultural differences.} \\
&Inclusive: The email acknowledges the different cultures &Inclusive: The email acknowledges the different cultures.\\
&Not inclusive: The email focuses on only one culture or does not acknowledges cultural differences.&Not inclusive: The email focuses on only one holiday, group, or culture and does not acknowledges cultural differences.\\
&Maybe: Not sure.&Maybe: Not sure. \\
\hline
Scale Item Generalized & \textit{Inclusive: The email emphasises an inclusive company culture by asking for participations from all kinds of traditions and cultural practices. It also uses inclusive language}&\textit{Inclusive: The email is focusing on all kinds of traditions and cultural practices. It also uses inclusive language}\\
\hline
%Option Definition Made More Specific & \textit{Inclusive: The email emphasises an inclusive company culture by asking for participations from all kinds of traditions and cultural practices. It also uses inclusive language}&\textit{Inclusive: The email is focusing on all kinds of traditions and cultural practices. It also uses inclusive language}\\
%&&Not Inclusive: The E-mail only promotes Christmas and does not ask for participations from other cultures and traditions
%&Not Inclusive: The email only promotes Christmas and does not ask for participations from other cultures and traditions\\

Minor Edit &\textit{What is the best summary?}&\textit{Which is the best summary?}\\
\hline
Instruction to Model  &\textit{Which of the following summaries best describe the article. The summary should be reflective of the key points in the article.} &\textit{Which of the following summaries best describe the article. The summary should be reflective of the key points in the article. Each article must be reviewed on its own and ignore all other summaries while doing so.}\\
\hline
Exclusion Criteria Added &\textit{The summary should be accurate and concise}&\textit{The summary should be accurate and concise. Has no fake data generated outside of reference}\\
\bottomrule
\end{tabular}
\caption{Types of changes made across every evaluation.}
\label{tab:changesmade}
\end{table*}

\begin{table*}[h!]
\small
\centering
\begin{tabular}{p{1.5cm}p{3.5cm}lp{6.5cm}}
\toprule
\textbf{Category}& \textbf{Definition}  & \textbf{Taxonomy}&\textbf{Example }\\ \midrule
 \textbf{Specificity}& The criteria ranges in how specific it is to the nature of the task context example & High Specificity& \textit{Please evaluate whether the following E-Mail is inclusive. This means that not only western traditions, such as Christmas, are celebrated, but employees are actively asked to contribute their customs and traditions to contribute to a diverse and inclusive company culture. Please also assess whether inclusive language is being used throughout the E-Mail.}\\
 &&Low Specificity&\textit{The email is inclusive.}\\
 \textbf{Additional Prompting} &The criteria includes instructions beyond criteria description& Present&\textit{Which summary offers the most clear and correct answer. As well as answering anything the customer did not know what to ask for in relation to their question. You must review each summary independently of the other summaries when making your judgment.} \\
 &&Absent&\textit{Which is the best summary?}\\
 \textbf{Rules}&If/else rules provided & Present& \textit{"Does the response refer to the  Travel and Expense policy? - Good
Makes exceptions to known rules (e.g., trip length different or different meeting days) - Bad
Does the response say all charges should be to corporate card while picking up on the fact that upgrading/business class is a personal expense....correctly dissociates the corporate from the personal expense -Good
Including rationale or policy statements - Good"}\\
 & & Absent & \textit{Is the response faithful according to the reference document?}\\
\textbf{Exclusion Criteria}&Criteria includes what should not be considered & Present &\textit{'The summary does not repeat unimportant details. }\\
&&Absent&\textit{The summary covers important aspects from the text.}\\
\textbf{Examples Provided}& Examples provided within criteria &Present& \textit{Criteria: The email should be inclusive taking into consideration all employees, 'Yes': "Happy Holidays! let's bring some gifts for all", 'No': "Merry Christmas! let's have Christmas tress!"} \\ 
&&Absent&\textit{Criteria: Is the email inclusive?,  'Yes': 'The email is inclusive.', 'No': 'The email is not inclusive.'}\\
\bottomrule
\end{tabular}
\caption{Categories of final criteria generated by participants for each task.}
\label{tab:criteriacode}
\end{table*}

\begin{table*}[h!]
\small
\centering
\begin{tabular}{p{2cm}p{5cm}p{7cm}}
\toprule
\textbf{Task}& \textbf{Interpretation of Criteria } & \textbf{Example}\\ \midrule
Email Inclusivity & Email must mention all possible cultures & 	\textit{The generated email should be inclusive. It should mention the different cultures and holidays (such as Christmas, Hanukkah, Kwanzaa, Diwali, and others) . It should not be exclusive to one culture. The email should use inclusive terms such as holiday and festive as opposed to terms exclusive to one culture.} \\	
&Email must not mention any specific culture & \textit{"The email is very formal and does not mention any specific gender, race, ethnicity, culture based terms or words.
"}\\		
&Email must not favor one culture to another&\textit{Which email invitation is most inclusive? An email is inclusive if it does not favour one religion over others and if it does not imply that all employees belong to a certain group or celebrate a specific holiday. An inclusive email refrains from using denominational words.}\\
&Gift exchange budget mentioned in email should not exceed particular amount&  \textit{"The price mentioned falls within the \$40 budget limit."}\\ 
\midrule
Summarization Preference &Summary must be factual&\textit{The summarization contains factual statements from the article}\\
&Summary must include key points from the article &\textit{The summary accurately conveys the main points of the article. }\\
&Summary must be succinct&\textit{The summary length is approximately 20\% of the original article length.}\\
&Summary must be grammatically correct&\textit{Is the response grammatically, lexically, semantically, and syntactically correct and common, including punctuation accuracy?}\\ 
&Summary must include specific details (detailed by user) from summary&\textit{Summary described dress worn by Vivien Leigh with details of price and date.}\\
&Summary should be the ``best''&\textit{Which is the best summary?}\\
\midrule 
Q\&A Faithfulness& Response must be grounded/faithful to the reference document &\textit{``Based on the reference document, does the response answer the customer question correctly?''}\\ 
&Response should adhere to company policy&\textit{``Has the employee claimed only the regular "in policy" fare and not the personal cost?''}\\ 
&Response should not include hallucinations&\textit{``Are there some hallucinations in the answer?''}\\
&Response should be correct&\textit{``Which summary offers the most clear and correct answer.''}\\
\bottomrule
\end{tabular}

\caption{Variations of interpretations of criteria within each task.}
\label{tab:interpretations}
\end{table*}

\subsection{RQ2: How do task-related factors and judge strategy impact how practitioners refine criteria?}
\begin{figure*}[htbp]
    \centering
    \begin{minipage}[b]{0.45\textwidth}
        \centering
        \includegraphics[width=\textwidth]{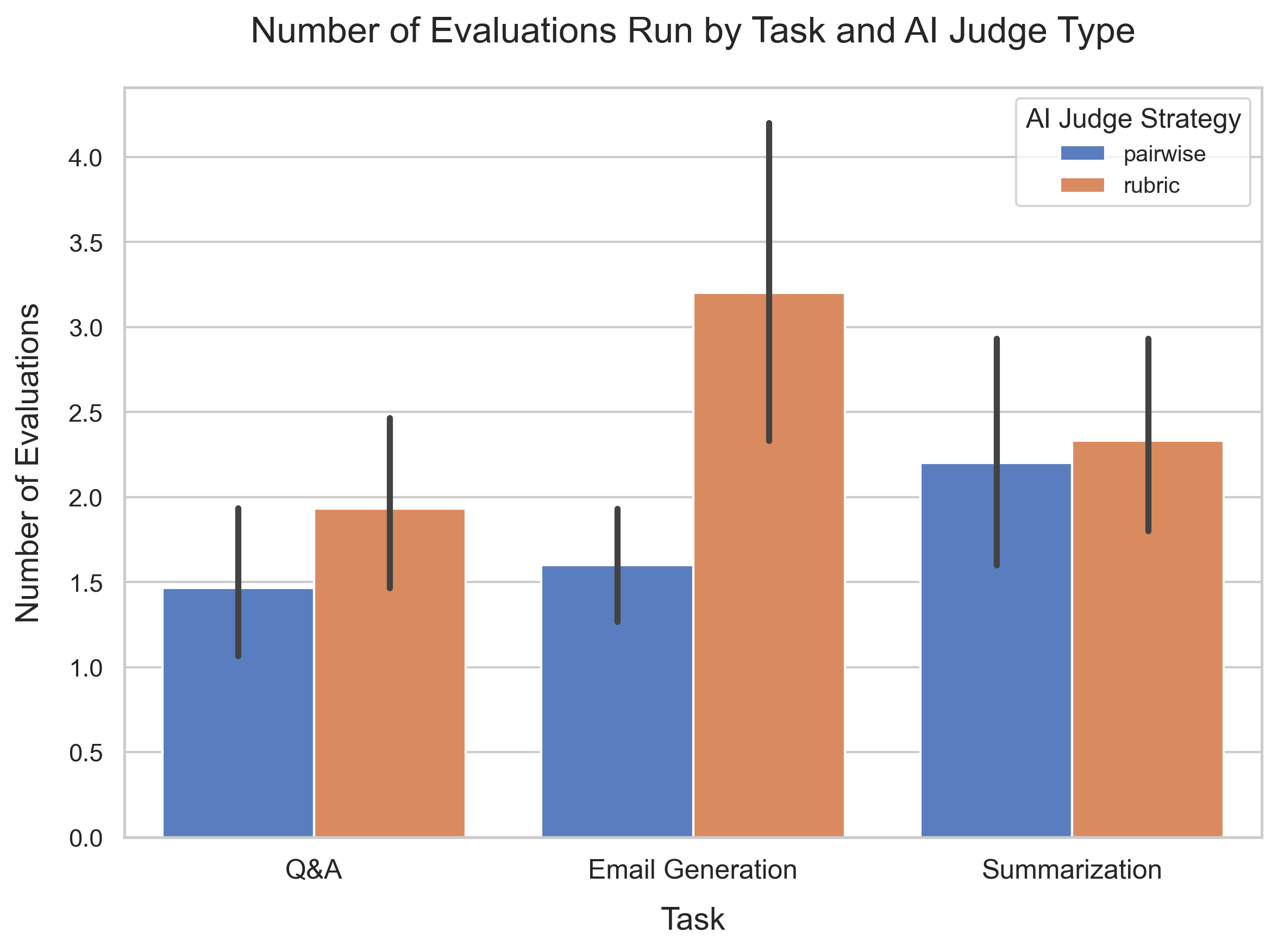}
        \caption{Participants conducted more evaluations in the direct assessment task compared to the pairwise task (p < 0.05).}
        \label{fig:evaluations}
    \end{minipage}
    \hfill
    \begin{minipage}[b]{0.45\textwidth}
        \centering
        \includegraphics[width=\textwidth]{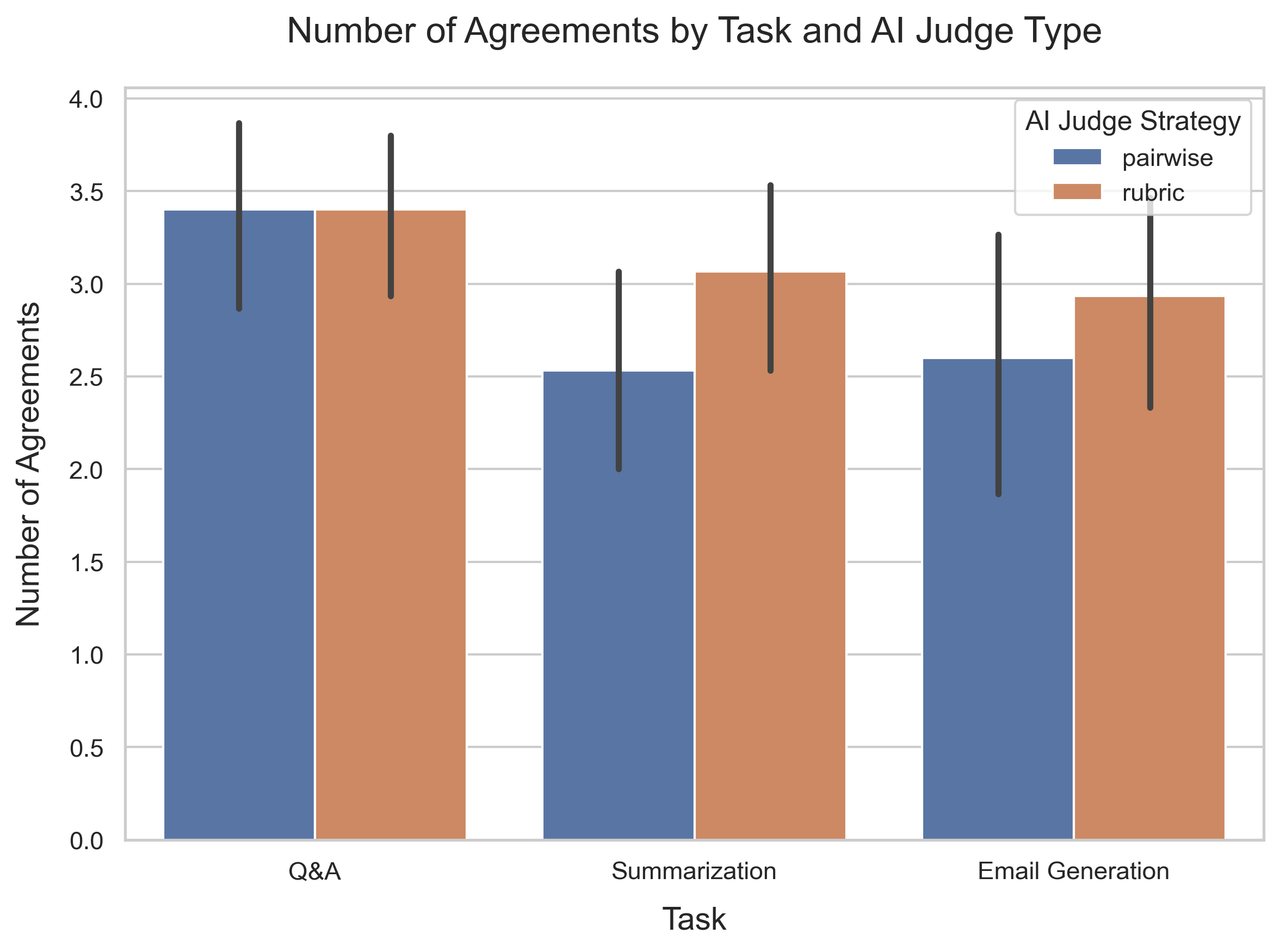}
        \caption{There were marginal significant differences in the degree of alignment achieved across task types (p = 0.08).}
        \label{fig:agreement}
    \end{minipage}
\end{figure*}

\subsubsection{What are the differences for the total amount of evaluations run?}
Participants were able to run multiple evaluations for each task yielding a total of 131 evaluations. We investigated whether there were differences in the strategies users employed when refining their criteria, particularly focusing on the total number of evaluations conducted. A repeated measures ANOVA was performed to compare the number of evaluations across different judge strategies (pairwise vs. rubric) and tasks (Q\&A Faithfulness, Email Inclusivity, Summarization). The results indicated a significant difference in the number of evaluations between judge strategies, with more evaluations being conducted in the direct assessment (rubric) condition compared to the pairwise condition. Specifically, the ANOVA showed that the judge strategy had a significant impact on the number of evaluations run (F(1, 14) = 12.64, p = 0.003), and the interaction between task and strategy was not significant. Post-hoc analysis using Tukey's HSD further confirmed that the pairwise strategy led to significantly fewer evaluations than the rubric strategy (mean difference = 0.72, p = 0.007). The number of evaluations can be seen in Figure \ref{fig:evaluations}

\subsubsection{What are the differences in human-AI alignment?}
We examined the the extent to which users agreed with the AI model's assessments across different conditions. Upon completion of each task, users indicated agreement with the model’s judgment. We conducted a repeated measures ANOVA, focusing on the judge strategies (pairwise vs. rubric) and the tasks (Q\&A Faithfulness, Email Inclusivity, Summarization). The analysis revealed a marginally significant main effect for task type, \(F(2, 28) = 2.73, p = 0.08\), indicating that there was a difference between alignment achieved between the tasks. Tukey post hoc analysis further confirmed this finding, showing a marginally significant difference in alignment between the the Q\&A task and the Email Generation Task (mean difference = -0.633, p = 0.08). The number of agreements can be seen in Figure \ref{fig:agreement}.

%faithfulness   inclusivity  -0.6333 0.0897 -1.3423 0.0756  Fal

%diparticipants aligned more with the AI model’s responses under the direct assessment (rubric) approach compared to the pairwise method. Tukey post hoc analysis further confirmed this finding, showing a significant difference in alignment between the pairwise and rubric strategies (mean difference = 2.67, p < 0.05). The number of agreements can be seen in Figure \ref{fig:agreement}. 

%The analysis revealed a significant main effect for judge strategy, \(F(1, 13) = 232.96, p < 0.001\), indicating that participants aligned more with the AI model’s responses under the direct assessment (rubric) approach compared to the pairwise method. Tukey post hoc analysis further confirmed this finding, showing a significant difference in alignment between the pairwise and rubric strategies (mean difference = 2.67, p < 0.05). The number of agreements can be seen in Figure \ref{fig:agreement}. 

\subsection{RQ3: How Do Task-Related Factors and Judge Strategy Impact User Perceptions?}

\begin{figure*}[htbp]
    \centering
    \begin{minipage}[b]{0.48\textwidth}
         \centering
        \includegraphics[width=\textwidth]{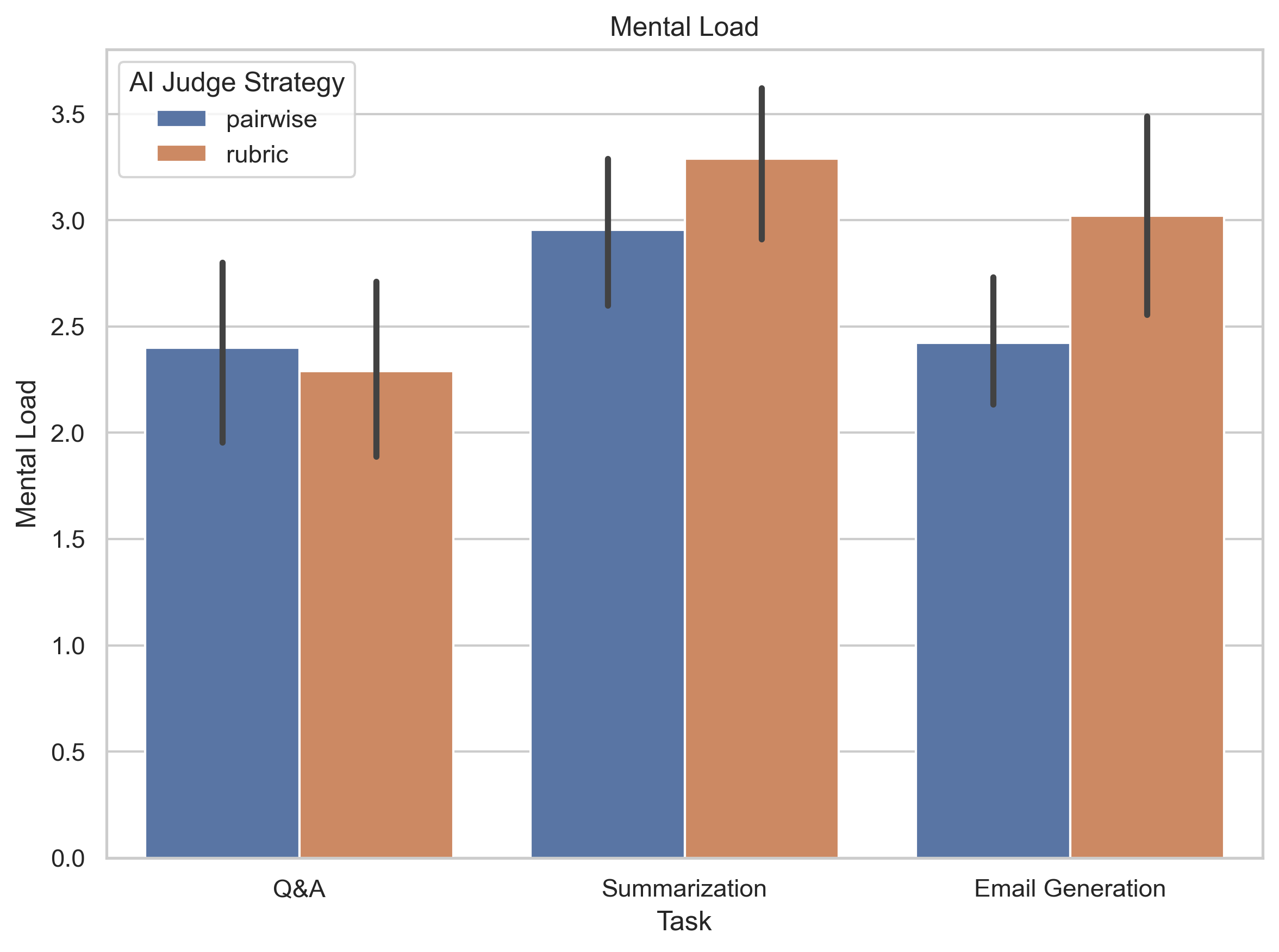}
        \caption{Users experienced higher cognitive load in the Summarization task than the Q\&A task p<0.05).}
        \label{fig:mentalload}
    \end{minipage}
    \hfill
    \begin{minipage}[b]{0.48\textwidth}
        \centering
        \includegraphics[width=\textwidth]{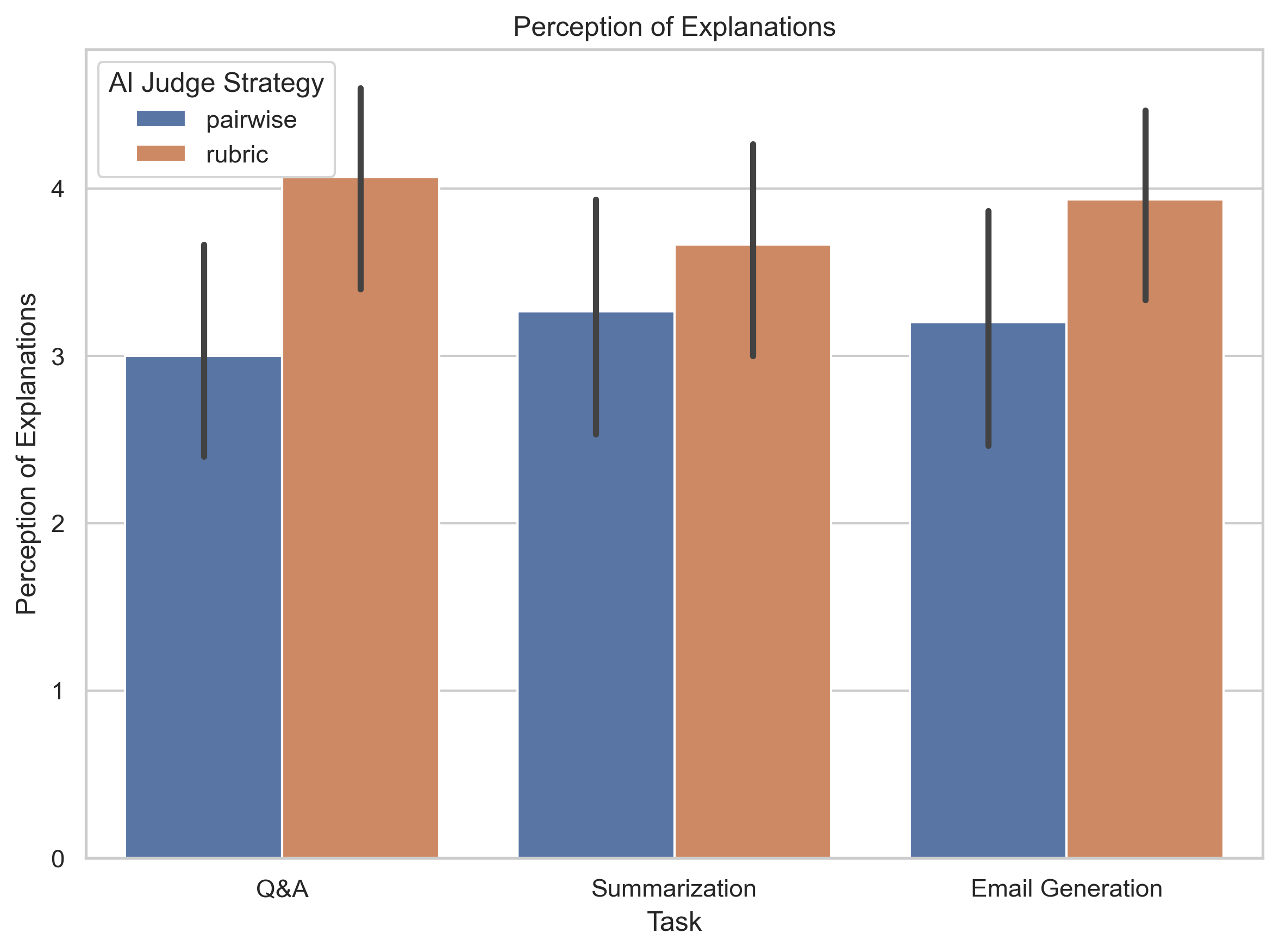}
        \caption{Explanations were perceived as more helpful in the rubric condition than in the pairwise condition (p<0.05). }
        \label{fig:exp}
    \end{minipage}

    \vspace{0.3cm} 

    % Second row with two images
   
\end{figure*}

We investigate\za{d} whether task-related factors and judge strategy impact user perceptions, specifically focusing on AI trust, perception of explanations, and cognitive load. For each of these measures, we conducted a repeated measures ANOVA, and we report the results below.

\subsubsection{AI Trust}
For trust in the AI evaluator, a repeated measures ANOVA showed no significant differences between tasks or AI strategies. Despite this, we identified reasons why participants either trusted or distrusted the AI evaluator. Participants who reported higher trust (above the median score of 4) highlighted several factors, such as the alignment between AI judgments and their criteria, prior positive experiences with specific models, and cohesive explanations. Some examples of feedback include:

\begin{quote} \textit{I didn't trust fully in the beginning, but the explanations helped.} \#P10 (Summarization, Rubric) \end{quote}

\begin{quote} \textit{Assessment aligned with my own.} \#P5 (Q\&A, Rubric) \end{quote}

\begin{quote} \textit{The AI judge was more accurate than me on my own faithfulness metric. It helped me identify gaps in my quick assessment.} \#P9 (Q\&A, Rubric) \end{quote}

\subsubsection{Reasons for Distrust}
Conversely, participants who expressed lower trust often felt that the AI model imposed its own criteria or failed to align with their expectations:

\begin{quote} \textit{The model was opinionated on the way to evaluate and didn’t enforce my criteria.} \#P4 (Email Generation, Pairwise) \end{quote}

%\subsubsection{Self-Satisfaction}

\subsubsection{Perception of Explanations}
We conducted an ANOVA to determine whether the type of  AI Judge Strategy (pairwise vs. rubric) affected participants' perception of how helpful the explanations were in completing tasks. The results indicated a significant main effect of the type of AI strategy, \( F(1, 14) = 7.79, p = 0.014 \), with rubric-based explanations rated more favorably overall compared to pairwise. However, there were no significant effects of task type, \( F(2, 28) = 0.09, p = 0.9152 \), nor was there a significant interaction between task and strategy, \( F(2, 28) = 0.67, p = 0.521 \).

To further explore these differences, we performed a post hoc analysis using Tukey’s HSD test. The results revealed that participants rated rubric explanations significantly higher than pairwise explanations (mean difference = 0.73, \( p = 0.008 \)), suggesting that rubric explanations were perceived as more useful across all tasks (Figure \ref{fig:exp}). We also analyzed participants' feedback on when explanations were perceived as helpful versus unhelpful. Explanations were deemed helpful in two key ways: they served as tools for refining criteria or as a means of validating existing criteria.

\begin{quote} \textit{After looking at the explanations, I realized that criteria can be improved and be made more specific. For the criteria I chose, I think I agree with the explanations provided, and it helps me understand the ranking better.} \#P10 (Email Generation, Pairwise) \end{quote}

Conversely, explanations were considered unhelpful when they were either not used or were found to be ineffective in resolving discrepancies.

\begin{quote} \textit{Explanations helped me see the shortcoming of AI, but the shortcomings were not able to be fixed with multiple iterations of criteria refinement.} \#P4 (Email Generation, Pairwise) \end{quote}

\subsubsection{Perception of Positional Bias} Of the total evaluations run by the 15 users, positional bias appeared in 35\% of those evaluations. We conducted an ANOVA to determine  whether the type of AI Judge strategy (pairwise vs. rubric) and task type (Q\&A, email generation, summarization) affected participants' perception of how helpful indication of positional bias was in completing the tasks. The analysis revealed no significant main effects, indicating that neither the type of task nor the strategy used by the judge had a significant impact on participants' perceptions of the helpfulness of positional bias. 
This lack of effect could potentially be explained by the fact that positional bias was only present in 35\% percent of evaluations. An independent samples t-test was conducted to compare the perceived helpfulness when participants results included positional bias and when they did not. The results showed a significant difference, t(28) = 2.98, p = 0.003, showing that when positional bias was present, participants reported it being more helpful than it did not appear in the results. The primary reason for positional bias not being helpful was that it did not appear in the results.

\begin{quote}
\textit{I quickly glanced to see if there was positional bias detected but it was not. Hence, I did not follow up on that. } \#P10 (Q\&A, Pairwise)
\end{quote}

Participants reported that the positional bias was helpful because it alerted them to rephrase their criteria.

\begin{quote}
\textit{Positional bias inspired me to simplify the criteria and rephrase them.} \#P2 (Summarization, Rubric)
\end{quote}

\begin{quote}
\textit{It alerted me that some rephrasing might be needed before expanding to the whole dataset.} \#P2 (Q\&A, Pairwise) 
\end{quote}

\subsubsection{Cognitive Load}
We also assessed cognitive load using three items from the NASA TLX (Task Load Index). The repeated measures ANOVA revealed a significant main effect of the task on cognitive load, \(F(2, 28) = 9.41, p = 0.0008\), indicating that the cognitive load varied significantly depending on the task. Post hoc analysis using Tukey's HSD test showed that cognitive load was significantly higher in the summarization task compared to the Q\&A task (meandiff = .779, p = 0.001) (see Figure \ref{fig:mentalload}). There was no significant effect of judge strategy on cognitive load, \(F(1, 14) = 3.02, p = 0.1044\), and the interaction between task and judge strategy also did not reach significance, \(F(2, 28) = 2.23, p = 0.126\).

\subsection{RQ4: What do practitioners prefer: Direct Assessment or Pairwise Comparison?}
Participants were asked to reflect on their preferred judgment strategy, whether direct assessment, pairwise comparison, or a context-dependent approach, and to explain the reasons behind their preference. Participant responses ranged with almost half (7 participants) reporting that they preferred direct assessment, and the remaining being split between pairwise and saying it depends/its contextual. The overview of preference reasons can be seen in Table \ref{tab:prefreasons}. Below we report why participants preferred one judge strategy over another.

\subsubsection{Preferred Direct Assessment}
Participants felt that direct assessment gave them more control over how the model outputs were evaluated and appreciated the detailed evaluation of the responses.

\begin{quote}
\textit{Direct assessment is more clear task for me to understand the results. Sometime ranks provided by pairwise assessment can be set in different order and still be correct.} \#P1
\end{quote}

 \begin{quote}
\textit{I prefer to create direct assessments because I feel that I can control more the outcomes of the model.} \#P14
\end{quote}

\begin{quote}
\textit{I would prefer the direct assessment mainly as it involves evaluating each response individually. Rankings are just relative preferences and they do not offer a detailed way of evaluating the model response. As an example in if we have all bad outputs, we would still have some ordering among them but that is essentially irrelevant to judge the response of the model which is not the case in rubric based criteria.} \#P3
\end{quote}

\subsubsection{Preferred Pairwise Assessment}
Of the 15 participants in the study, 4 reported preferring pairwise assessment over direct assessment. The reasons included that participants felt that pairwise assessment allowed for more flexibility and nuance in evaluating subjective tasks  where a binary or rigid criteria may not apply. Others reported that the pairwise was simpler and easier to user when formulating evaluation criteria, since detailed rubrics may be more complex. 

\begin{quote}
\textit{"Pairwise is easier to use for crafting an evaluation criteria, especially for the summarization task where the 'goodness' of a summary is very abstract and can be broad."} \#P7
\end{quote}
\begin{quote}
\textit{It's easier to formulate a sentence containing the complete criteria and not have to think about rubrics.} \#P5
\end{quote}

\subsubsection{Use of AI Judge Strategy is Contextual}
Participants who preferred neither judge strategy  reported that they would either like to use a combination of strategies to accomplish their task, or that their preference would depend on the nature of the task. 

\begin{quote}
\textit{Various tasks are of various complexities, I feel that ranking and classification results coming out of the pairwise and direct assessment can help the user make better decisions which are both relative and independent.} \#P4
\end{quote}

\begin{quote}
\textit{I believe they have different use cases. Rubric can be used in a case of classification or compliance for example. I might want to consider all answers and see what classifications I can assign to them. Pairwise is about getting the best result for what you want. I don't care as much about the others, only that one is the correct.} \#P8
\end{quote}

\section{Limitations}
\zauist{While this study offers insights into how practitioners refine evaluation criteria and interact with AI-assisted judgment strategies, there are a few limitations. First, participants were recruited from a single organization. Although they had relevant domain expertise, this recruitment context may limit the generalizability of our findings. Future work should expand to more diverse populations, including external practitioners with varied evaluation needs and workflows. Additionally, even though \texttt{EvalAssist} was designed with scalability in mind through features such as task abstraction and flexible input configurations, its capabilities were only explored in small-scale, interactive scenarios in this study. Future work should test criteria generalization across larger datasets, and varied prompts.  Third, the study relied primarily on participants’ self-reported perceptions of trust, cognitive load, and explanation quality. While subjective in nature, these responses offer valuable insight into users’ real-time judgments and experiences during evaluation. Explanations were also more accessible in the Direct Assessment condition than in the Pairwise condition, which may have influenced trust and perception. Future work can examine how explanation visibility shapes user trust and evaluation behavior.} 

\section{Discussion}

%Pos bias indicators neither increased nor decreased user trust 
%\textcolor{red}{include quantitative results here}
%- no trust differences ... why? 
%- explanations more helpful in the rubric condition 
%- explanations were more 
%and people did more in direct assessment 
%MORE CONTROL 

\subsection{Variability in Subjective Criteria: Defining Stakeholder Needs in AI-assisted Evaluation}
We observed significant variation in how participants defined criteria within tasks, a finding consistent with prior HCI research on diversity in subjective judgments \cite{karapanos2009accounting}. For example, "inclusivity" had different interpretations among participants. Some participants believed it meant mentioning every possible holiday to represent all religions, while others thought it meant avoiding specific cultural or holiday references altogether. This aligns with studies in inclusive design, where users with diverse backgrounds interpret fairness and inclusivity differently, highlighting the challenges of capturing subjective values in AI systems \cite{costanza2020design}. This variation underscores the importance of clear stakeholder definitions for criteria definitions, a need echoed in crowdsourcing research, which shows that subjective tasks can lead to inconsistent evaluations without precise guidelines \cite{kittur2013future}. When developing criteria to evaluate model outputs, it is essential to define stakeholder needs clearly, especially for subjective criteria like inclusivity.  

\za{The tasks assigned to users (email generation, summarization preference, and Q\&A faithfulness), differed in their levels of specificity and ambiguity. However, we observed that regardless of how specific the instructions were, people consistently interpreted the task in significantly different ways. Ultimately, it is the human who defines the evaluation criteria and communicates these expectations to the AI-evaluator.
These findings highlight the importance of careful deliberation and decision-making with all stakeholders and impacted parties when defining evaluation criteria. For example, stakeholders must explicitly decide what terms like "inclusive" or "a good summary" mean before relying on an LLM judge. Adopting deliberative processes like participatory design with diverse stakeholders, including end-users,  can contribute to ensure multiple perspectives are considered. These steps ensure that the resulting criteria is aligned with the needs of all parties involved. Additionally, differences in alignment were observed across task types, suggesting that some tasks may naturally achieve better alignment than others. This underscores the need for machine learning practitioners to work with all relevant stakeholders to agree on clear, precise directions tailored to the specific task to ensure consistent and effective evaluations.}
%\textcolor{red}{differences in tasks - write a paragraph then }
%\textcolor{red}{add a bit about summarization and differences seen in sumamrizatiom? }
%\textcolor{red}{To address your comment, we have added more discussion about the nature of the tasks and the observed differences, particularly how these relate to RQ2. This provides a clearer explanation of our approach and findings.}

\subsubsection{Mitigating Over-Specificity in AI-Assisted Evaluation: Balancing Task Context and Generalization}
\label{sec:specific}
When analyzing how participants adjusted their criteria to align with the AI's judgments, we observed a consistent tendency to make the criteria increasingly specific to the task at hand. This resulted in overly narrow criteria that worked well for evaluating a single article's summary but failed to generalize across multiple summaries from different articles. Other AI-assisted evaluation systems \cite{kim2023evallm,shankar2024validates} do not necessarily prioritize sampling diverse task contexts and outputs, especially when the evaluation heavily depends on prompts. This limits the user's exposure to varied outputs, increasing the risk of overspecifying criteria to a particular task. To address this issue, it is critical to expose users to diverse task contexts. For instance, in the case of article summarization, this would mean providing users with varied articles and summaries to encourage the development of criteria that can be applied more broadly. A design improvement could involve enabling ML practitioners to upload large datasets, with the system recommending or allowing the selection of diverse samples to help develop more robust criteria.

Conversely, we found that a subset of users kept constructs very general. For example they would rely on the AI evaluator to interpret what it means to be ``inclusive'' or what it means to be grounded in the document. This lack of specificity can also lead to subpar results. Striking a balance between overly specific and overly vague criteria is essential. Encouraging users to refine their criteria while keeping them general enough for broad application can improve quality. Paired with exposure to diverse outputs, this approach can create more effective evaluation processes that work well across different contexts. \zauist{Some participants over-specified their criteria by embedding ground-truth expectations, while others generalized too broadly. As previously discussed, the nature of the representative data influenced how participants refined their criteria. These observations show the importance of selecting diverse and representative examples, including multiple instances per task, to support effective and generalizable criteria refinement. This finding directly informs our design recommendations for improving LLM-as-a-judge systems.}

\begin{table*}[htbp]
\centering
\begin{tabular}{@{}p{2.8cm}p{4.7cm}p{8.2cm}@{}}
\toprule
\textbf{Preference} & \textbf{Code} & \textbf{Definition} \\
\midrule
\multirow{2}{*}{Direct Assessment} 
    & Clarity and Control & Offers a clearer, more understandable evaluation process with greater user control. \\
    & Detailed Evaluation & Enables response-by-response evaluation, which is helpful for tasks requiring thorough review. \\
\midrule
\multirow{2}{*}{Pairwise Comparison} 
    & Flexibility and Nuance & Supports nuanced evaluation of abstract or subjective tasks. \\
    & Ease of Criteria Formulation & Simplifies the creation of evaluation criteria when rubrics are difficult to define. \\
\midrule
\multirow{2}{*}{Neither (Contextual)} 
    & Combination Approach & A combined use of both methods is effective for complex tasks, balancing classification and ranking needs. \\
    & Task-Specific Fit & Direct Assessment is suited for compliance/classification, while Pairwise is better for preference-based selection. \\
\bottomrule
\end{tabular}
\caption{Codes and rationales for users’ preferred evaluation strategies.}
\label{tab:prefreasons}
\end{table*}

\subsubsection{Challenges of Natural Language Criteria Formulation} 
\label{sec:transparency}
When users expressed their criteria in natural language, we observed significant variation in how they structured their inputs, particularly in the pairwise condition where no predefined form or structure was provided. Some participants introduced additional instructions, asking the AI evaluator to consider each response independently, which contradicts the purpose of pairwise comparisons where outputs are meant to be directly compared. Others created complex rule-based systems, such as: \textit{If the response includes [insert characteristic], then it is ``good''; if it does not, then it is ``bad''.} Additionally, rather than defining item scales as instructed, participants often provided examples of "good" and "bad" responses. These alternative interaction patterns did not align with the intended functionality of the evaluation process and led to suboptimal results. To address this, greater transparency should be provided regarding the prompts used in both direct and pairwise evaluation conditions, ensuring users have a clearer understanding of how their input will be processed. Providing more examples that users can customize to fit their specific needs could also improve outcomes, as well as implementing auto-correction features to guide users in entering the criteria correctly.

\begin{figure}
    \centering
    \includegraphics[width=.8\linewidth]{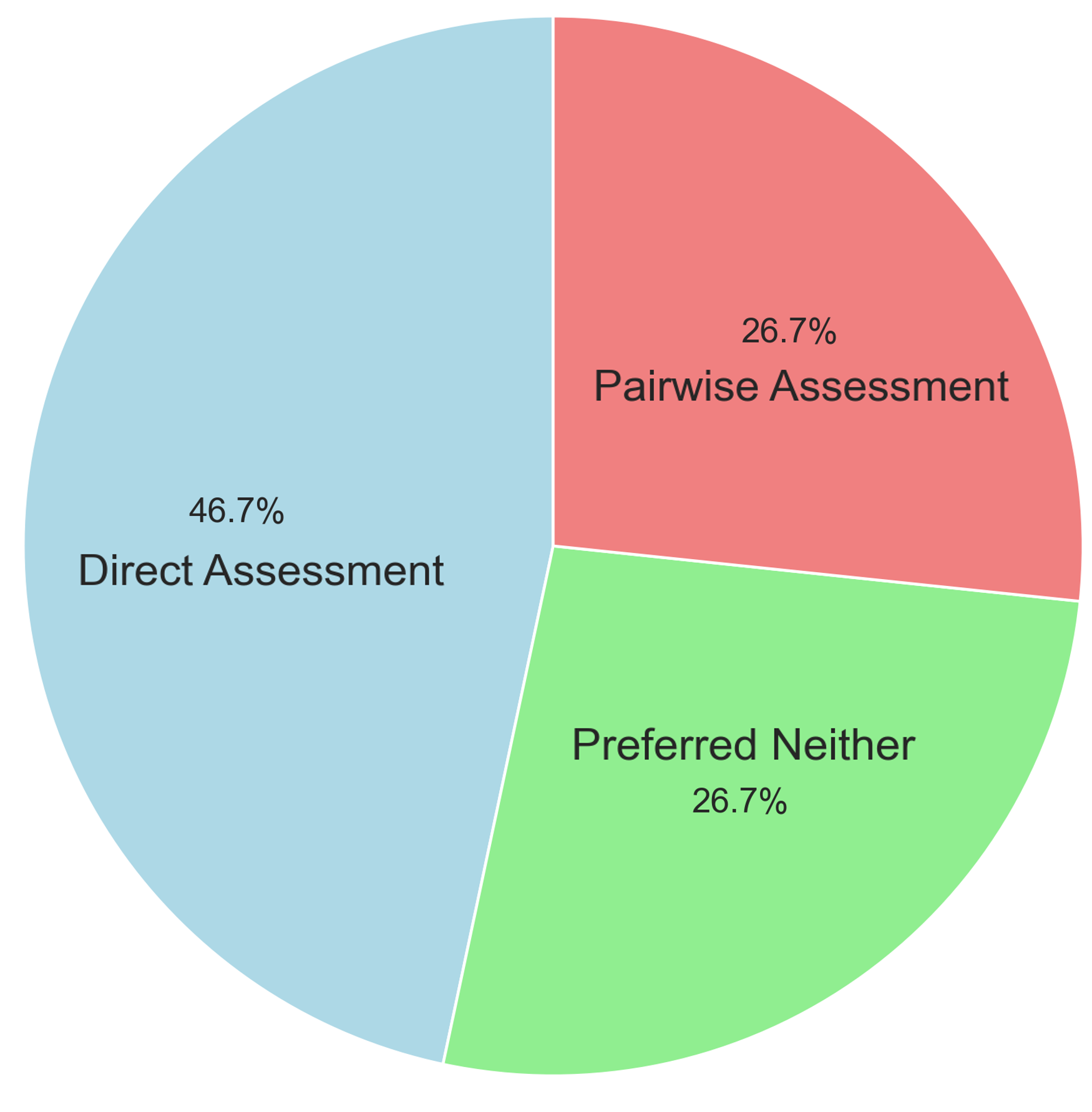}
    \caption{Assessment Preferences}
    \label{fig:pie}
\end{figure}

\subsection{Alignment: Refining Criteria, Changing the AI Evaluator, and Altering Expected Results}
We observed various user interactions throughout the evaluation process that mirror principles seen in interactive machine learning (iML), where users iteratively refine models by selecting examples, labeling data, and evaluating outputs \cite{patel2010gestalt, simard2014ice}. In line with iML approaches, users in our study adjusted their criteria throughout the evaluation, making them more general or specific as needed. This adaptive behavior reflects the need for flexible interfaces that allow seamless transitions between specifying, revising, and evaluating criteria. As noted in iML, reducing user effort while improving output alignment is crucial \cite{desmond2022ai}. In some cases, midway through the evaluation, users even changed their expected outcomes to align more closely with the AI's judgments, especially after reading the explanations, which often convinced them of the AI's output. Users frequently revised their criteria, whether by adding or removing scales in the rubric condition, or rephrasing their criteria in the pairwise condition. They also experimented with different AI evaluators, switching between models like Mixtral and LLaMA to compare results. This suggests that the path to "alignment" can take many forms, from adjusting one's own expectations, to switching the AI model making the judgments, to modifying the criteria itself in search of a better fit. This process could be better supported in the future by automatically testing criteria with multiple models and showing results side-by-side. \za{Future work can investigate the impact of different types of refinement on alignment and, if significant patterns emerge, explore tooling that nudges users toward specific refinements to improve alignment and evaluation outcomes.}

%Also, a recommendation system could help guide users in refining their criteria. 

\subsection{Adaptive Evaluation Strategies: Balancing Clarity and Flexibility}
Participants expressed a preference for direct assessment due to the clarity and control it offered, allowing for detailed, individualized evaluations. On the other hand, pairwise comparison was valued for its flexibility, particularly in more subjective tasks. Some participants preferred a hybrid approach, selecting an evaluation method based on the specific task. The higher number of evaluations in direct assessment suggests that users felt more engaged, likely because it provided them with a greater sense of control. This aligns with the reasons participants cited for favoring direct assessment. These findings highlight the importance of adaptable evaluation strategies tailored to task type. AI systems could benefit from offering multiple evaluation modes or an adaptive judgment strategy that adjusts based on task complexity. For example, users might opt for direct assessment for objective tasks and switch to pairwise ranking for more subjective ones.
Additionally, some users might prefer a combination approach, starting with direct assessment to filter out responses and then using pairwise ranking to further refine the best options. Future research can explore hybrid systems that adapt dynamically based on user workflows.

%\za{add something here about the number of high evaluations }

\subsection{\za{Trust Across Judgment Strategies}}
\za{We observed no significant differences in trust across task types or between the two AI judge strategies (direct assessment and pairwise comparison). This indicates that participants did not inherently favor one strategy over the other in terms of trustworthiness, and the design of these strategies did not introduce unintended trust disparities. While prior literature highlights how task framing and algorithmic presentation can disproportionately influence trust \cite{greiner2024incentives}, our findings suggest that neither judge strategy nor task type significantly impacted trust. Open-ended feedback revealed another area for exploration: the influence of bias indicators and explanations on trust. Participants noted that both bias and the quality of explanations played a role in shaping their trust in the AI evaluator. Explanations were particularly valued in the rubric condition, suggesting that future research could investigate how different explanation types affect trust across contexts. Overall, it is encouraging that no major trust differences were observed between pairwise and direct assessment strategies, indicating both are perceived as equally trustworthy by users.}

\subsection{Bias Awareness and Explanation Visibility in Evaluation Strategies} 
Positional bias was considered helpful when present, prompting participants to revise and improve their criteria. This result suggests that highlighting AI biases can enhance the evaluation process and improve alignment between human and AI judgments. While our study focused on positional bias, there is an opportunity to explore the representation of other biases, such as self-enhancement bias and verbosity bias. For example, is the model ranking outputs highly because they were also generated by the AI evaluator LLM, or is it favoring longer responses even if they are less accurate or clear? \za{ Incorporating a suite of bias indicators tailored to different contexts and tasks could strengthen users’ ability to evaluate outputs effectively. This approach aligns with the goal of fostering transparency and user trust in AI systems by equipping users with the tools to identify and mitigate potential biases.}

Explanations were perceived as more helpful in the direct assessment condition. One reason, as reported by participants, is that in the pairwise condition, explanations often went unnoticed. Since every comparison generates an explanation, these explanations were only accessible via a modal pop-up, requiring users to click on the result. Even then, users had to sift through multiple explanations to understand the ranking. This limitation in the pairwise condition presents an opportunity to redesign how explanations are presented, making them more concise and digestible. When explanations were more visible, as in the direct assessment condition, they proved to be more effective, suggesting a need for improvement in how pairwise explanations are displayed.

%\subsection{Future directions }

\section{Conclusion}
We introduced \texttt{EvalAssist} a tool designed to help practitioners refine evaluation criteria using both direct and pairwise judgment strategies. The tool provides positional bias metrics and explanations for each AI judgment. In a controlled experiment with 15 machine learning practitioners, we examined how users refine their criteria and identified key differences between the two evaluation approaches. Direct assessment was preferred for its clarity and control, while pairwise comparison was valued for flexibility in subjective tasks.  Users often refined their criteria by increasing specificity or simplifying scales to improve alignment with task needs. Human-AI alignment was stronger in the rubric-based (direct assessment) condition, highlighting the importance of clear evaluation criteria.  While our participants found \texttt{EvalAssist} a valuable tool to be integrated into their work practices, findings from our experiment highlight areas for future improvements: (1) AI-assisted evaluation systems should offer flexible, adaptive strategies, allowing users to switch between direct and pairwise methods, (2) they should support ongoing criteria refinement, provide clear explanations, and improve alignment between human and AI judgments to enhance evaluation reliability, and (3) they should support involving multiple stakeholders in the criteria refinement process to pave the way to alignment. %Inspired by our user research, we are currently in the process of rolling out an evolved AI-assisted evaluation tool to a larger user population to observe "usage in the wild."

\bibliographystyle{ACM-Reference-Format}
\bibliography{sample}


\begin{thebibliography}{56}

%%% the Web address.

\ifx \showCODEN    \undefined \def \showCODEN     #1{\unskip}     \fi
\ifx \showISBNx    \undefined \def \showISBNx     #1{\unskip}     \fi
\ifx \showISBNxiii \undefined \def \showISBNxiii  #1{\unskip}     \fi
\ifx \showISSN     \undefined \def \showISSN      #1{\unskip}     \fi
\ifx \showLCCN     \undefined \def \showLCCN      #1{\unskip}     \fi
\ifx \shownote     \undefined \def \shownote      #1{#1}          \fi
\ifx \showarticletitle \undefined \def \showarticletitle #1{#1}   \fi
\ifx \showURL      \undefined \def \showURL       {\relax}        \fi
% invisible to TeX
\providecommand\bibfield[2]{#2}
\providecommand\bibinfo[2]{#2}
\providecommand\natexlab[1]{#1}
\providecommand\showeprint[2][]{arXiv:#2}

\bibitem[Anthropic(2024)]%
        {anthropic2024claude35}
\bibfield{author}{\bibinfo{person}{Anthropic}.} \bibinfo{year}{2024}\natexlab{}.
\newblock \bibinfo{title}{Claude 3.5 sonnet}.
\newblock
\urldef\tempurl%
\url{https://www.anthropic.com/news/claude-3-5-sonnet}
\showURL{%
\tempurl}
\newblock
\shownote{Accessed: 2024-09-09}.


\bibitem[Arawjo et~al\mbox{.}(2024)]%
        {arawjo2024chainforge}
\bibfield{author}{\bibinfo{person}{Ian Arawjo}, \bibinfo{person}{Chelse Swoopes}, \bibinfo{person}{Priyan Vaithilingam}, \bibinfo{person}{Martin Wattenberg}, {and} \bibinfo{person}{Elena~L Glassman}.} \bibinfo{year}{2024}\natexlab{}.
\newblock \bibinfo{title}{ChainForge: A Visual Toolkit for Prompt Engineering and LLM Hypothesis Testing}.
\newblock \bibinfo{numpages}{18}~pages.
\newblock


\bibitem[Bai et~al\mbox{.}(2024)]%
        {BaiYingCao2024}
\bibfield{author}{\bibinfo{person}{Yushi Bai}, \bibinfo{person}{Jiahao Ying}, \bibinfo{person}{Yixin Cao}, \bibinfo{person}{Xin Lv}, \bibinfo{person}{Yuze He}, \bibinfo{person}{Xiaozhi Wang}, \bibinfo{person}{Jifan Yu}, \bibinfo{person}{Kaisheng Zeng}, \bibinfo{person}{Yijia Xiao}, \bibinfo{person}{Haozhe Lyu}, \bibinfo{person}{Jiayin Zhang}, \bibinfo{person}{Juanzi Li}, {and} \bibinfo{person}{Lei Hou}.} \bibinfo{year}{2024}\natexlab{}.
\newblock \bibinfo{title}{Benchmarking foundation models with language-model-as-an-examiner}.
\newblock \bibinfo{numpages}{26}~pages.
\newblock


\bibitem[Bavaresco et~al\mbox{.}(2024)]%
        {Bavaresco2024JUDGE_BENCH}
\bibfield{author}{\bibinfo{person}{Anna Bavaresco}, \bibinfo{person}{Raffaella Bernardi}, \bibinfo{person}{Leonardo Bertolazzi}, \bibinfo{person}{Desmond Elliott}, \bibinfo{person}{Raquel Fernández}, \bibinfo{person}{Albert Gatt}, \bibinfo{person}{E. Ghaleb}, \bibinfo{person}{Mario Giulianelli}, \bibinfo{person}{Michael Hanna}, \bibinfo{person}{Alexander Koller}, \bibinfo{person}{André F.~T. Martins}, \bibinfo{person}{Philipp Mondorf}, \bibinfo{person}{Vera Neplenbroek}, \bibinfo{person}{Sandro Pezzelle}, \bibinfo{person}{Barbara Plank}, \bibinfo{person}{David Schlangen}, \bibinfo{person}{Alessandro Suglia}, \bibinfo{person}{Aditya~K Surikuchi}, \bibinfo{person}{Ece Takmaz}, {and} \bibinfo{person}{Alberto Testoni}.} \bibinfo{year}{2024}\natexlab{}.
\newblock \bibinfo{title}{LLMs instead of Human Judges? A Large Scale Empirical Study across 20 NLP Evaluation Tasks}.
\newblock
\urldef\tempurl%
\url{https://arxiv.org/abs/2406.18403}
\showURL{%
\tempurl}


\bibitem[Brachman et~al\mbox{.}(2024)]%
        {brachman2024knowledge}
\bibfield{author}{\bibinfo{person}{Michelle Brachman}, \bibinfo{person}{Amina El-Ashry}, \bibinfo{person}{Casey Dugan}, {and} \bibinfo{person}{Werner Geyer}.} \bibinfo{year}{2024}\natexlab{}.
\newblock \bibinfo{title}{How Knowledge Workers Use and Want to Use LLMs in an Enterprise Context}.
\newblock \bibinfo{numpages}{8}~pages.
\newblock


\bibitem[Bu{\c{c}}inca et~al\mbox{.}(2020)]%
        {buccinca2020proxy}
\bibfield{author}{\bibinfo{person}{Zana Bu{\c{c}}inca}, \bibinfo{person}{Phoebe Lin}, \bibinfo{person}{Krzysztof~Z Gajos}, {and} \bibinfo{person}{Elena~L Glassman}.} \bibinfo{year}{2020}\natexlab{}.
\newblock \bibinfo{title}{Proxy tasks and subjective measures can be misleading in evaluating explainable AI systems}.
\newblock \bibinfo{numpages}{454--464}~pages.
\newblock


\bibitem[Chen et~al\mbox{.}(2024)]%
        {chen2024humans}
\bibfield{author}{\bibinfo{person}{Guiming~Hardy Chen}, \bibinfo{person}{Shunian Chen}, \bibinfo{person}{Ziche Liu}, \bibinfo{person}{Feng Jiang}, {and} \bibinfo{person}{Benyou Wang}.} \bibinfo{year}{2024}\natexlab{}.
\newblock \bibinfo{title}{Humans or llms as the judge? a study on judgement biases}.
\newblock


\bibitem[Chiang and Lee(2023)]%
        {chiang-lee-2023-closer}
\bibfield{author}{\bibinfo{person}{Cheng-Han Chiang} {and} \bibinfo{person}{Hung-yi Lee}.} \bibinfo{year}{2023}\natexlab{}.
\newblock \showarticletitle{A Closer Look into Using Large Language Models for Automatic Evaluation}. In \bibinfo{booktitle}{\emph{Findings of the Association for Computational Linguistics: EMNLP 2023}}, \bibfield{editor}{\bibinfo{person}{Houda Bouamor}, \bibinfo{person}{Juan Pino}, {and} \bibinfo{person}{Kalika Bali}} (Eds.). \bibinfo{publisher}{Association for Computational Linguistics}, \bibinfo{address}{Singapore}, \bibinfo{pages}{8928--8942}.
\newblock
\href{https://doi.org/10.18653/v1/2023.findings-emnlp.599}{doi:\nolinkurl{10.18653/v1/2023.findings-emnlp.599}}


\bibitem[Chiang and yi~Lee(2023)]%
        {chiang2023can}
\bibfield{author}{\bibinfo{person}{Cheng-Han Chiang} {and} \bibinfo{person}{Hung yi Lee}.} \bibinfo{year}{2023}\natexlab{}.
\newblock \showarticletitle{Can Large Language Models Be an Alternative to Human Evaluations?}. In \bibinfo{booktitle}{\emph{Proceedings of the 61st Annual Meeting of the Association for Computational Linguistics (Volume 1: Long Papers)}}. \bibinfo{publisher}{Association for Computational Linguistics}, \bibinfo{address}{Toronto, Canada}, \bibinfo{pages}{15607--15631}.
\newblock


\bibitem[Chung et~al\mbox{.}(2022)]%
        {https://doi.org/10.48550/arxiv.2210.11416}
\bibfield{author}{\bibinfo{person}{Hyung~Won Chung}, \bibinfo{person}{Le Hou}, \bibinfo{person}{Shayne Longpre}, \bibinfo{person}{Barret Zoph}, \bibinfo{person}{Yi Tay}, \bibinfo{person}{William Fedus}, \bibinfo{person}{Eric Li}, \bibinfo{person}{Xuezhi Wang}, \bibinfo{person}{Mostafa Dehghani}, \bibinfo{person}{Siddhartha Brahma}, \bibinfo{person}{Albert Webson}, \bibinfo{person}{Shixiang~Shane Gu}, \bibinfo{person}{Zhuyun Dai}, \bibinfo{person}{Mirac Suzgun}, \bibinfo{person}{Xinyun Chen}, \bibinfo{person}{Aakanksha Chowdhery}, \bibinfo{person}{Sharan Narang}, \bibinfo{person}{Gaurav Mishra}, \bibinfo{person}{Adams Yu}, \bibinfo{person}{Vincent Zhao}, \bibinfo{person}{Yanping Huang}, \bibinfo{person}{Andrew Dai}, \bibinfo{person}{Hongkun Yu}, \bibinfo{person}{Slav Petrov}, \bibinfo{person}{Ed~H. Chi}, \bibinfo{person}{Jeff Dean}, \bibinfo{person}{Jacob Devlin}, \bibinfo{person}{Adam Roberts}, \bibinfo{person}{Denny Zhou}, \bibinfo{person}{Quoc~V. Le}, {and} \bibinfo{person}{Jason Wei}.}
  \bibinfo{year}{2022}\natexlab{}.
\newblock \bibinfo{title}{Scaling Instruction-Finetuned Language Models}.
\newblock
\href{https://doi.org/10.48550/ARXIV.2210.11416}{doi:\nolinkurl{10.48550/ARXIV.2210.11416}}


\bibitem[Costanza-Chock(2020)]%
        {costanza2020design}
\bibfield{author}{\bibinfo{person}{Sasha Costanza-Chock}.} \bibinfo{year}{2020}\natexlab{}.
\newblock \bibinfo{title}{Design justice: Community-led practices to build the worlds we need}.
\newblock


\bibitem[Desmond et~al\mbox{.}(2024)]%
        {desmond2024evalullm}
\bibfield{author}{\bibinfo{person}{Michael Desmond}, \bibinfo{person}{Zahra Ashktorab}, \bibinfo{person}{Qian Pan}, \bibinfo{person}{Casey Dugan}, {and} \bibinfo{person}{James~M. Johnson}.} \bibinfo{year}{2024}\natexlab{}.
\newblock \showarticletitle{EvaluLLM: LLM assisted evaluation of generative outputs}. In \bibinfo{booktitle}{\emph{Companion Proceedings of the 29th International Conference on Intelligent User Interfaces}} (Greenville, SC, USA) \emph{(\bibinfo{series}{IUI '24 Companion})}. \bibinfo{publisher}{Association for Computing Machinery}, \bibinfo{address}{New York, NY, USA}, \bibinfo{pages}{30–32}.
\newblock
\showISBNx{9798400705090}
\href{https://doi.org/10.1145/3640544.3645216}{doi:\nolinkurl{10.1145/3640544.3645216}}


\bibitem[Desmond and Brachman(2024)]%
        {desmond2024exploring}
\bibfield{author}{\bibinfo{person}{Michael Desmond} {and} \bibinfo{person}{Michelle Brachman}.} \bibinfo{year}{2024}\natexlab{}.
\newblock \bibinfo{title}{Exploring Prompt Engineering Practices in the Enterprise}.
\newblock


\bibitem[Desmond et~al\mbox{.}(2022)]%
        {desmond2022ai}
\bibfield{author}{\bibinfo{person}{Michael Desmond}, \bibinfo{person}{Michelle Brachman}, \bibinfo{person}{Evelyn Duesterwald}, \bibinfo{person}{Casey Dugan}, \bibinfo{person}{Narendra~Nath Joshi}, \bibinfo{person}{Qian Pan}, {and} \bibinfo{person}{Carolina Spina}.} \bibinfo{year}{2022}\natexlab{}.
\newblock \bibinfo{title}{AI Assisted Data Labeling with Interactive Auto Label}.
\newblock \bibinfo{numpages}{13161--13163}~pages.
\newblock


\bibitem[Dubois et~al\mbox{.}(2024)]%
        {dubois2024alpacafarm}
\bibfield{author}{\bibinfo{person}{Yann Dubois}, \bibinfo{person}{Chen~Xuechen Li}, \bibinfo{person}{Rohan Taori}, \bibinfo{person}{Tianyi Zhang}, \bibinfo{person}{Ishaan Gulrajani}, \bibinfo{person}{Jimmy Ba}, \bibinfo{person}{Carlos Guestrin}, \bibinfo{person}{Percy~S Liang}, {and} \bibinfo{person}{Tatsunori~B Hashimoto}.} \bibinfo{year}{2024}\natexlab{}.
\newblock \bibinfo{title}{Alpacafarm: A simulation framework for methods that learn from human feedback}.
\newblock


\bibitem[Fabbri et~al\mbox{.}(2021)]%
        {fabbri2021summeval}
\bibfield{author}{\bibinfo{person}{Alexander~R Fabbri}, \bibinfo{person}{Wojciech Kry{\'s}ci{\'n}ski}, \bibinfo{person}{Bryan McCann}, \bibinfo{person}{Caiming Xiong}, \bibinfo{person}{Richard Socher}, {and} \bibinfo{person}{Dragomir Radev}.} \bibinfo{year}{2021}\natexlab{}.
\newblock \bibinfo{title}{Summeval: Re-evaluating summarization evaluation}.
\newblock \bibinfo{numpages}{391--409}~pages.
\newblock


\bibitem[Floridi and Chiriatti(2020)]%
        {floridi2020gpt}
\bibfield{author}{\bibinfo{person}{Luciano Floridi} {and} \bibinfo{person}{Massimo Chiriatti}.} \bibinfo{year}{2020}\natexlab{}.
\newblock \bibinfo{title}{GPT-3: Its nature, scope, limits, and consequences}.
\newblock \bibinfo{numpages}{681--694}~pages.
\newblock


\bibitem[for Research~on Foundation~Models(2024)]%
        {helm}
\bibfield{author}{\bibinfo{person}{Center for Research~on Foundation~Models}.} \bibinfo{year}{2024}\natexlab{}.
\newblock \bibinfo{title}{HELM: Holistic Evaluation of Language Models}.
\newblock \bibinfo{howpublished}{\url{https://crfm.stanford.edu/helm/}}.
\newblock
\newblock
\shownote{Accessed: 2024-11-20}.


\bibitem[Gao et~al\mbox{.}(2023)]%
        {gao2023retrieval}
\bibfield{author}{\bibinfo{person}{Yunfan Gao}, \bibinfo{person}{Yun Xiong}, \bibinfo{person}{Xinyu Gao}, \bibinfo{person}{Kangxiang Jia}, \bibinfo{person}{Jinliu Pan}, \bibinfo{person}{Yuxi Bi}, \bibinfo{person}{Yi Dai}, \bibinfo{person}{Jiawei Sun}, {and} \bibinfo{person}{Haofen Wang}.} \bibinfo{year}{2023}\natexlab{}.
\newblock \bibinfo{title}{Retrieval-augmented generation for large language models: A survey}.
\newblock


\bibitem[Gehrmann et~al\mbox{.}(2023)]%
        {gehrmann2023repairing}
\bibfield{author}{\bibinfo{person}{Sebastian Gehrmann}, \bibinfo{person}{Elizabeth Clark}, {and} \bibinfo{person}{Thibault Sellam}.} \bibinfo{year}{2023}\natexlab{}.
\newblock \bibinfo{title}{Repairing the cracked foundation: A survey of obstacles in evaluation practices for generated text}.
\newblock \bibinfo{numpages}{103--166}~pages.
\newblock


\bibitem[Ghazal et~al\mbox{.}(2013)]%
        {ghazal2013bigbench}
\bibfield{author}{\bibinfo{person}{Ahmad Ghazal}, \bibinfo{person}{Tilmann Rabl}, \bibinfo{person}{Minqing Hu}, \bibinfo{person}{Francois Raab}, \bibinfo{person}{Meikel Poess}, \bibinfo{person}{Alain Crolotte}, {and} \bibinfo{person}{Hans-Arno Jacobsen}.} \bibinfo{year}{2013}\natexlab{}.
\newblock \bibinfo{title}{Bigbench: Towards an industry standard benchmark for big data analytics}.
\newblock \bibinfo{numpages}{1197--1208}~pages.
\newblock


\bibitem[Greiner et~al\mbox{.}(2024)]%
        {greiner2024incentives}
\bibfield{author}{\bibinfo{person}{Ben Greiner}, \bibinfo{person}{Philipp Gr{\"u}nwald}, \bibinfo{person}{Thomas Lindner}, \bibinfo{person}{Georg Lintner}, {and} \bibinfo{person}{Martin Wiernsperger}.} \bibinfo{year}{2024}\natexlab{}.
\newblock \bibinfo{title}{Incentives, Framing, and Reliance on Algorithmic Advice: An Experimental Study}.
\newblock


\bibitem[Hart(2006)]%
        {hart2006nasa}
\bibfield{author}{\bibinfo{person}{Sandra~G Hart}.} \bibinfo{year}{2006}\natexlab{}.
\newblock \bibinfo{title}{NASA-task load index (NASA-TLX); 20 years later}.
\newblock \bibinfo{numpages}{904--908}~pages.
\newblock


\bibitem[Hendrycks et~al\mbox{.}(2020)]%
        {hendrycks2020measuring}
\bibfield{author}{\bibinfo{person}{Dan Hendrycks}, \bibinfo{person}{Collin Burns}, \bibinfo{person}{Steven Basart}, \bibinfo{person}{Andy Zou}, \bibinfo{person}{Mantas Mazeika}, \bibinfo{person}{Dawn Song}, {and} \bibinfo{person}{Jacob Steinhardt}.} \bibinfo{year}{2020}\natexlab{}.
\newblock \bibinfo{title}{Measuring massive multitask language understanding}.
\newblock


\bibitem[Huang et~al\mbox{.}(2024)]%
        {huang2024empirical}
\bibfield{author}{\bibinfo{person}{Hui Huang}, \bibinfo{person}{Yingqi Qu}, \bibinfo{person}{Jing Liu}, \bibinfo{person}{Muyun Yang}, {and} \bibinfo{person}{Tiejun Zhao}.} \bibinfo{year}{2024}\natexlab{}.
\newblock \bibinfo{title}{An empirical study of llm-as-a-judge for llm evaluation: Fine-tuned judge models are task-specific classifiers}.
\newblock


\bibitem[Jiang et~al\mbox{.}(2023)]%
        {jiang2023mistral}
\bibfield{author}{\bibinfo{person}{Albert~Q Jiang}, \bibinfo{person}{Alexandre Sablayrolles}, \bibinfo{person}{Arthur Mensch}, \bibinfo{person}{Chris Bamford}, \bibinfo{person}{Devendra~Singh Chaplot}, \bibinfo{person}{Diego de~las Casas}, \bibinfo{person}{Florian Bressand}, \bibinfo{person}{Gianna Lengyel}, \bibinfo{person}{Guillaume Lample}, \bibinfo{person}{Lucile Saulnier}, {et~al\mbox{.}}} \bibinfo{year}{2023}\natexlab{}.
\newblock \bibinfo{title}{Mistral 7B}.
\newblock


\bibitem[Kahng et~al\mbox{.}(2024)]%
        {kahng2024llm}
\bibfield{author}{\bibinfo{person}{Minsuk Kahng}, \bibinfo{person}{Ian Tenney}, \bibinfo{person}{Mahima Pushkarna}, \bibinfo{person}{Michael~Xieyang Liu}, \bibinfo{person}{James Wexler}, \bibinfo{person}{Emily Reif}, \bibinfo{person}{Krystal Kallarackal}, \bibinfo{person}{Minsuk Chang}, \bibinfo{person}{Michael Terry}, {and} \bibinfo{person}{Lucas Dixon}.} \bibinfo{year}{2024}\natexlab{}.
\newblock \bibinfo{title}{Llm comparator: Visual analytics for side-by-side evaluation of large language models}.
\newblock \bibinfo{numpages}{7}~pages.
\newblock


\bibitem[Karapanos et~al\mbox{.}(2009)]%
        {karapanos2009accounting}
\bibfield{author}{\bibinfo{person}{Evangelos Karapanos}, \bibinfo{person}{Jean-Bernard Martens}, {and} \bibinfo{person}{Marc Hassenzahl}.} \bibinfo{year}{2009}\natexlab{}.
\newblock \bibinfo{title}{Accounting for diversity in subjective judgments}.
\newblock \bibinfo{numpages}{639--648}~pages.
\newblock


\bibitem[Kim et~al\mbox{.}(2023)]%
        {kim2023prometheus}
\bibfield{author}{\bibinfo{person}{Seungone Kim}, \bibinfo{person}{Jamin Shin}, \bibinfo{person}{Yejin Cho}, \bibinfo{person}{Joel Jang}, \bibinfo{person}{Shayne Longpre}, \bibinfo{person}{Hwaran Lee}, \bibinfo{person}{Sangdoo Yun}, \bibinfo{person}{Seongjin Shin}, \bibinfo{person}{Sungdong Kim}, \bibinfo{person}{James Thorne}, {et~al\mbox{.}}} \bibinfo{year}{2023}\natexlab{}.
\newblock \bibinfo{title}{Prometheus: Inducing fine-grained evaluation capability in language models}.
\newblock


\bibitem[Kim et~al\mbox{.}(2024b)]%
        {kim2024prometheus}
\bibfield{author}{\bibinfo{person}{Seungone Kim}, \bibinfo{person}{Juyoung Suk}, \bibinfo{person}{Shayne Longpre}, \bibinfo{person}{Bill~Yuchen Lin}, \bibinfo{person}{Jamin Shin}, \bibinfo{person}{Sean Welleck}, \bibinfo{person}{Graham Neubig}, \bibinfo{person}{Moontae Lee}, \bibinfo{person}{Kyungjae Lee}, {and} \bibinfo{person}{Minjoon Seo}.} \bibinfo{year}{2024}\natexlab{b}.
\newblock \bibinfo{title}{Prometheus 2: An open source language model specialized in evaluating other language models}.
\newblock


\bibitem[Kim et~al\mbox{.}(2024a)]%
        {kim2023evallm}
\bibfield{author}{\bibinfo{person}{Tae~Soo Kim}, \bibinfo{person}{Yoonjoo Lee}, \bibinfo{person}{Jamin Shin}, \bibinfo{person}{Young-Ho Kim}, {and} \bibinfo{person}{Juho Kim}.} \bibinfo{year}{2024}\natexlab{a}.
\newblock \showarticletitle{EvalLM: Interactive Evaluation of Large Language Model Prompts on User-Defined Criteria}. In \bibinfo{booktitle}{\emph{Proceedings of the 2024 CHI Conference on Human Factors in Computing Systems}} (Honolulu, HI, USA) \emph{(\bibinfo{series}{CHI '24})}. \bibinfo{publisher}{Association for Computing Machinery}, \bibinfo{address}{New York, NY, USA}, Article \bibinfo{articleno}{306}, \bibinfo{numpages}{21}~pages.
\newblock
\showISBNx{9798400703300}
\href{https://doi.org/10.1145/3613904.3642216}{doi:\nolinkurl{10.1145/3613904.3642216}}


\bibitem[Kittur et~al\mbox{.}(2013)]%
        {kittur2013future}
\bibfield{author}{\bibinfo{person}{Aniket Kittur}, \bibinfo{person}{Jeffrey~V Nickerson}, \bibinfo{person}{Michael Bernstein}, \bibinfo{person}{Elizabeth Gerber}, \bibinfo{person}{Aaron Shaw}, \bibinfo{person}{John Zimmerman}, \bibinfo{person}{Matt Lease}, {and} \bibinfo{person}{John Horton}.} \bibinfo{year}{2013}\natexlab{}.
\newblock \bibinfo{title}{The future of crowd work}.
\newblock \bibinfo{numpages}{1301--1318}~pages.
\newblock


\bibitem[Kryscinski et~al\mbox{.}(2020)]%
        {kryscinski-etal-2020-evaluating}
\bibfield{author}{\bibinfo{person}{Wojciech Kryscinski}, \bibinfo{person}{Bryan McCann}, \bibinfo{person}{Caiming Xiong}, {and} \bibinfo{person}{Richard Socher}.} \bibinfo{year}{2020}\natexlab{}.
\newblock \showarticletitle{Evaluating the Factual Consistency of Abstractive Text Summarization}. In \bibinfo{booktitle}{\emph{Proceedings of the 2020 Conference on Empirical Methods in Natural Language Processing (EMNLP)}}, \bibfield{editor}{\bibinfo{person}{Bonnie Webber}, \bibinfo{person}{Trevor Cohn}, \bibinfo{person}{Yulan He}, {and} \bibinfo{person}{Yang Liu}} (Eds.). \bibinfo{publisher}{Association for Computational Linguistics}, \bibinfo{address}{Online}, \bibinfo{pages}{9332--9346}.
\newblock
\href{https://doi.org/10.18653/v1/2020.emnlp-main.750}{doi:\nolinkurl{10.18653/v1/2020.emnlp-main.750}}


\bibitem[Lewis et~al\mbox{.}(2020)]%
        {lewis2020retrieval}
\bibfield{author}{\bibinfo{person}{Patrick Lewis}, \bibinfo{person}{Ethan Perez}, \bibinfo{person}{Aleksandra Piktus}, \bibinfo{person}{Fabio Petroni}, \bibinfo{person}{Vladimir Karpukhin}, \bibinfo{person}{Naman Goyal}, \bibinfo{person}{Heinrich K{\"u}ttler}, \bibinfo{person}{Mike Lewis}, \bibinfo{person}{Wen-tau Yih}, \bibinfo{person}{Tim Rockt{\"a}schel}, {et~al\mbox{.}}} \bibinfo{year}{2020}\natexlab{}.
\newblock \bibinfo{title}{Retrieval-augmented generation for knowledge-intensive nlp tasks}.
\newblock \bibinfo{numpages}{9459--9474}~pages.
\newblock


\bibitem[Li et~al\mbox{.}(2019)]%
        {li2019acuteevalimproveddialogueevaluation}
\bibfield{author}{\bibinfo{person}{Margaret Li}, \bibinfo{person}{Jason Weston}, {and} \bibinfo{person}{Stephen Roller}.} \bibinfo{year}{2019}\natexlab{}.
\newblock \bibinfo{title}{ACUTE-EVAL: Improved Dialogue Evaluation with Optimized Questions and Multi-turn Comparisons}.
\newblock
\showeprint[arxiv]{1909.03087}~[cs.CL]
\urldef\tempurl%
\url{https://arxiv.org/abs/1909.03087}
\showURL{%
\tempurl}


\bibitem[Li et~al\mbox{.}(2023)]%
        {li2023alpacaeval}
\bibfield{author}{\bibinfo{person}{Xuechen Li}, \bibinfo{person}{Tianyi Zhang}, \bibinfo{person}{Yann Dubois}, \bibinfo{person}{Rohan Taori}, \bibinfo{person}{Ishaan Gulrajani}, \bibinfo{person}{Carlos Guestrin}, \bibinfo{person}{Percy Liang}, {and} \bibinfo{person}{Tatsunori~B Hashimoto}.} \bibinfo{year}{2023}\natexlab{}.
\newblock \bibinfo{title}{Alpacaeval: An automatic evaluator of instruction-following models}.
\newblock


\bibitem[Li et~al\mbox{.}(2024)]%
        {li2024split}
\bibfield{author}{\bibinfo{person}{Zongjie Li}, \bibinfo{person}{Chaozheng Wang}, \bibinfo{person}{Pingchuan Ma}, \bibinfo{person}{Daoyuan Wu}, \bibinfo{person}{Shuai Wang}, \bibinfo{person}{Cuiyun Gao}, {and} \bibinfo{person}{Yang Liu}.} \bibinfo{year}{2024}\natexlab{}.
\newblock \bibinfo{title}{Split and Merge: Aligning Position Biases in LLM-based Evaluators}.
\newblock \bibinfo{numpages}{11084--11108}~pages.
\newblock


\bibitem[Liang et~al\mbox{.}(2022)]%
        {liang2022holistic}
\bibfield{author}{\bibinfo{person}{Percy Liang}, \bibinfo{person}{Rishi Bommasani}, \bibinfo{person}{Tony Lee}, \bibinfo{person}{Dimitris Tsipras}, \bibinfo{person}{Dilara Soylu}, \bibinfo{person}{Michihiro Yasunaga}, \bibinfo{person}{Yian Zhang}, \bibinfo{person}{Deepak Narayanan}, \bibinfo{person}{Yuhuai Wu}, \bibinfo{person}{Ananya Kumar}, {et~al\mbox{.}}} \bibinfo{year}{2022}\natexlab{}.
\newblock \bibinfo{title}{Holistic evaluation of language models}.
\newblock


\bibitem[Liu et~al\mbox{.}(2023)]%
        {liu2023g}
\bibfield{author}{\bibinfo{person}{Yang Liu}, \bibinfo{person}{Dan Iter}, \bibinfo{person}{Yichong Xu}, \bibinfo{person}{Shuohang Wang}, \bibinfo{person}{Ruochen Xu}, {and} \bibinfo{person}{Chenguang Zhu}.} \bibinfo{year}{2023}\natexlab{}.
\newblock \bibinfo{title}{G-eval: Nlg evaluation using gpt-4 with better human alignment}.
\newblock


\bibitem[Liu et~al\mbox{.}(2024)]%
        {liu2024aligning}
\bibfield{author}{\bibinfo{person}{Yinhong Liu}, \bibinfo{person}{Han Zhou}, \bibinfo{person}{Zhijiang Guo}, \bibinfo{person}{Ehsan Shareghi}, \bibinfo{person}{Ivan Vuli{\'c}}, \bibinfo{person}{Anna Korhonen}, {and} \bibinfo{person}{Nigel Collier}.} \bibinfo{year}{2024}\natexlab{}.
\newblock \bibinfo{title}{Aligning with human judgement: The role of pairwise preference in large language model evaluators}.
\newblock


\bibitem[Madsen et~al\mbox{.}(2024)]%
        {madsen2024self}
\bibfield{author}{\bibinfo{person}{Andreas Madsen}, \bibinfo{person}{Sarath Chandar}, {and} \bibinfo{person}{Siva Reddy}.} \bibinfo{year}{2024}\natexlab{}.
\newblock \bibinfo{title}{Are self-explanations from Large Language Models faithful?}
\newblock \bibinfo{numpages}{295--337}~pages.
\newblock


\bibitem[Pan et~al\mbox{.}(2024)]%
        {pan2024human}
\bibfield{author}{\bibinfo{person}{Qian Pan}, \bibinfo{person}{Zahra Ashktorab}, \bibinfo{person}{Michael Desmond}, \bibinfo{person}{Martin~Santillan Cooper}, \bibinfo{person}{James Johnson}, \bibinfo{person}{Rahul Nair}, \bibinfo{person}{Elizabeth Daly}, {and} \bibinfo{person}{Werner Geyer}.} \bibinfo{year}{2024}\natexlab{}.
\newblock \bibinfo{title}{Human-Centered Design Recommendations for LLM-as-a-Judge}.
\newblock


\bibitem[Patel et~al\mbox{.}(2010)]%
        {patel2010gestalt}
\bibfield{author}{\bibinfo{person}{Kayur Patel}, \bibinfo{person}{Naomi Bancroft}, \bibinfo{person}{Steven~M Drucker}, \bibinfo{person}{James Fogarty}, \bibinfo{person}{Amy~J Ko}, {and} \bibinfo{person}{James Landay}.} \bibinfo{year}{2010}\natexlab{}.
\newblock \bibinfo{title}{Gestalt: integrated support for implementation and analysis in machine learning}.
\newblock \bibinfo{numpages}{37--46}~pages.
\newblock


\bibitem[Poursabzi-Sangdeh et~al\mbox{.}(2021)]%
        {poursabzi2021manipulating}
\bibfield{author}{\bibinfo{person}{Forough Poursabzi-Sangdeh}, \bibinfo{person}{Daniel~G Goldstein}, \bibinfo{person}{Jake~M Hofman}, \bibinfo{person}{Jennifer~Wortman Wortman~Vaughan}, {and} \bibinfo{person}{Hanna Wallach}.} \bibinfo{year}{2021}\natexlab{}.
\newblock \bibinfo{title}{Manipulating and measuring model interpretability}.
\newblock \bibinfo{numpages}{52}~pages.
\newblock


\bibitem[Quevedo et~al\mbox{.}(2025)]%
        {10.1007/978-3-031-86623-4_13}
\bibfield{author}{\bibinfo{person}{Ernesto Quevedo}, \bibinfo{person}{Jorge~Yero Salazar}, \bibinfo{person}{Rachel Koerner}, \bibinfo{person}{Pablo Rivas}, {and} \bibinfo{person}{Tomas Cerny}.} \bibinfo{year}{2025}\natexlab{}.
\newblock \showarticletitle{Detecting Hallucinations in Large Language Model Generation: A Token Probability Approach}. In \bibinfo{booktitle}{\emph{Artificial Intelligence and Applications}}, \bibfield{editor}{\bibinfo{person}{Hamid~R. Arabnia}, \bibinfo{person}{Leonidas Deligiannidis}, \bibinfo{person}{Soheyla Amirian}, \bibinfo{person}{Farzan Shenavarmasouleh}, \bibinfo{person}{Farid Ghareh~Mohammadi}, {and} \bibinfo{person}{David de~la Fuente}} (Eds.). \bibinfo{publisher}{Springer Nature Switzerland}, \bibinfo{address}{Cham}, \bibinfo{pages}{154--173}.
\newblock
\showISBNx{978-3-031-86623-4}


\bibitem[Raju et~al\mbox{.}(2024)]%
        {raju2024constructing}
\bibfield{author}{\bibinfo{person}{Ravi Raju}, \bibinfo{person}{Swayambhoo Jain}, \bibinfo{person}{Bo Li}, \bibinfo{person}{Jonathan Li}, {and} \bibinfo{person}{Urmish Thakkar}.} \bibinfo{year}{2024}\natexlab{}.
\newblock \bibinfo{title}{Constructing Domain-Specific Evaluation Sets for LLM-as-a-judge}.
\newblock


\bibitem[Reid et~al\mbox{.}(2024)]%
        {reid2024gemini}
\bibfield{author}{\bibinfo{person}{Machel Reid}, \bibinfo{person}{Nikolay Savinov}, \bibinfo{person}{Denis Teplyashin}, \bibinfo{person}{Dmitry Lepikhin}, \bibinfo{person}{Timothy Lillicrap}, \bibinfo{person}{Jean-baptiste Alayrac}, \bibinfo{person}{Radu Soricut}, \bibinfo{person}{Angeliki Lazaridou}, \bibinfo{person}{Orhan Firat}, \bibinfo{person}{Julian Schrittwieser}, {et~al\mbox{.}}} \bibinfo{year}{2024}\natexlab{}.
\newblock \bibinfo{title}{Gemini 1.5: Unlocking multimodal understanding across millions of tokens of context}.
\newblock


\bibitem[Saito et~al\mbox{.}(2023)]%
        {saito2023verbosity}
\bibfield{author}{\bibinfo{person}{Keita Saito}, \bibinfo{person}{Akifumi Wachi}, \bibinfo{person}{Koki Wataoka}, {and} \bibinfo{person}{Youhei Akimoto}.} \bibinfo{year}{2023}\natexlab{}.
\newblock \bibinfo{title}{Verbosity bias in preference labeling by large language models}.
\newblock


\bibitem[Shankar et~al\mbox{.}(2024)]%
        {shankar2024validates}
\bibfield{author}{\bibinfo{person}{Shreya Shankar}, \bibinfo{person}{JD Zamfirescu-Pereira}, \bibinfo{person}{Bj{\"o}rn Hartmann}, \bibinfo{person}{Aditya~G Parameswaran}, {and} \bibinfo{person}{Ian Arawjo}.} \bibinfo{year}{2024}\natexlab{}.
\newblock \bibinfo{title}{Who Validates the Validators? Aligning LLM-Assisted Evaluation of LLM Outputs with Human Preferences}.
\newblock


\bibitem[Shi et~al\mbox{.}(2024)]%
        {shi2024judging}
\bibfield{author}{\bibinfo{person}{Lin Shi}, \bibinfo{person}{Chiyu Ma}, \bibinfo{person}{Wenhua Liang}, \bibinfo{person}{Weicheng Ma}, {and} \bibinfo{person}{Soroush Vosoughi}.} \bibinfo{year}{2024}\natexlab{}.
\newblock \bibinfo{title}{Judging the Judges: A Systematic Investigation of Position Bias in Pairwise Comparative Assessments by LLMs}.
\newblock


\bibitem[Simard et~al\mbox{.}(2014)]%
        {simard2014ice}
\bibfield{author}{\bibinfo{person}{Patrice Simard}, \bibinfo{person}{David Chickering}, \bibinfo{person}{Aparna Lakshmiratan}, \bibinfo{person}{Denis Charles}, \bibinfo{person}{L{\'e}on Bottou}, \bibinfo{person}{Carlos Garcia~Jurado Suarez}, \bibinfo{person}{David Grangier}, \bibinfo{person}{Saleema Amershi}, \bibinfo{person}{Johan Verwey}, {and} \bibinfo{person}{Jina Suh}.} \bibinfo{year}{2014}\natexlab{}.
\newblock \bibinfo{title}{Ice: enabling non-experts to build models interactively for large-scale lopsided problems}.
\newblock


\bibitem[Thomas(2006)]%
        {thomas2006general}
\bibfield{author}{\bibinfo{person}{David~R Thomas}.} \bibinfo{year}{2006}\natexlab{}.
\newblock \bibinfo{title}{A general inductive approach for analyzing qualitative evaluation data}.
\newblock \bibinfo{numpages}{237--246}~pages.
\newblock


\bibitem[Wei et~al\mbox{.}(2022)]%
        {wei2022chain}
\bibfield{author}{\bibinfo{person}{Jason Wei}, \bibinfo{person}{Xuezhi Wang}, \bibinfo{person}{Dale Schuurmans}, \bibinfo{person}{Maarten Bosma}, \bibinfo{person}{Fei Xia}, \bibinfo{person}{Ed Chi}, \bibinfo{person}{Quoc~V Le}, \bibinfo{person}{Denny Zhou}, {et~al\mbox{.}}} \bibinfo{year}{2022}\natexlab{}.
\newblock \bibinfo{title}{Chain-of-thought prompting elicits reasoning in large language models}.
\newblock \bibinfo{numpages}{24824--24837}~pages.
\newblock


\bibitem[Xu et~al\mbox{.}(2024)]%
        {xu2024pride}
\bibfield{author}{\bibinfo{person}{Wenda Xu}, \bibinfo{person}{Guanglei Zhu}, \bibinfo{person}{Xuandong Zhao}, \bibinfo{person}{Liangming Pan}, \bibinfo{person}{Lei Li}, {and} \bibinfo{person}{William Wang}.} \bibinfo{year}{2024}\natexlab{}.
\newblock \bibinfo{title}{Pride and prejudice: LLM amplifies self-bias in self-refinement}.
\newblock \bibinfo{numpages}{15474--15492}~pages.
\newblock


\bibitem[Yu et~al\mbox{.}(2024)]%
        {yu2024evaluation}
\bibfield{author}{\bibinfo{person}{Hao Yu}, \bibinfo{person}{Aoran Gan}, \bibinfo{person}{Kai Zhang}, \bibinfo{person}{Shiwei Tong}, \bibinfo{person}{Qi Liu}, {and} \bibinfo{person}{Zhaofeng Liu}.} \bibinfo{year}{2024}\natexlab{}.
\newblock \bibinfo{title}{Evaluation of Retrieval-Augmented Generation: A Survey}.
\newblock


\bibitem[Zheng et~al\mbox{.}(2024)]%
        {ZhengChiangSheng2024}
\bibfield{author}{\bibinfo{person}{Lianmin Zheng}, \bibinfo{person}{Wei-Lin Chiang}, \bibinfo{person}{Ying Sheng}, \bibinfo{person}{Siyuan Zhuang}, \bibinfo{person}{Zhanghao Wu}, \bibinfo{person}{Yonghao Zhuang}, \bibinfo{person}{Zi Lin}, \bibinfo{person}{Zhuohan Li}, \bibinfo{person}{Dacheng Li}, \bibinfo{person}{Eric~P. Xing}, \bibinfo{person}{Hao Zhang}, \bibinfo{person}{Joseph~E. Gonzalez}, {and} \bibinfo{person}{Ion Stoica}.} \bibinfo{year}{2024}\natexlab{}.
\newblock \bibinfo{title}{Judging LLM-as-a-judge with MT-bench and Chatbot Arena}.
\newblock \bibinfo{numpages}{29}~pages.
\newblock


\end{thebibliography}

\appendix

\section{Additional Figures}

\begin{figure}[H]
    \centering
 \includegraphics[width=1\linewidth]{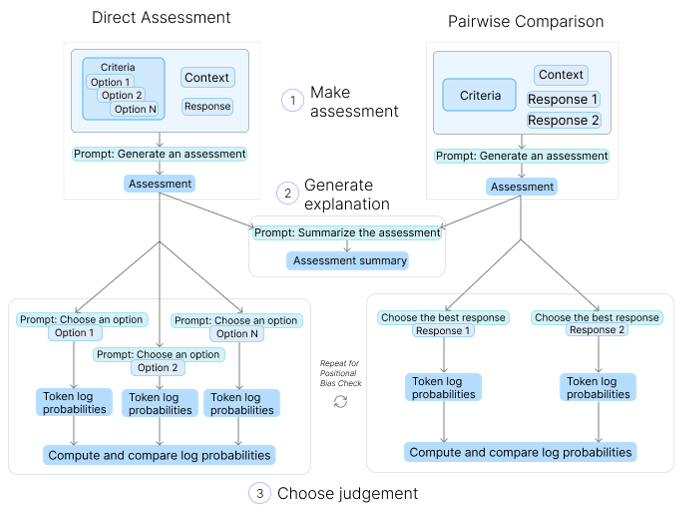}
    \caption{Generation of judgments in \texttt{EvalAssist} involves initially creating an assessment, followed by summarizing the assessment and making a selection between options. Log probabilities are generated and compared during this process, which is repeated to detect positional bias. The prompts shown have been simplified for clarity in the image.}
    \label{fig:evalassist}
\end{figure}

\section{Direct Assessment Prompts for EvalAssist AI Evaluators}
\label{appendix:prompts}

\subsection{Mixtral-8x7b-instruct-v01}
\subsubsection{Assessment Prompt}
\noindent\rule{\textwidth}{0pt} % Adds a horizontal line spanning the text width
\texttt{<s> [INST]}
\texttt{You are presented with a response generated subject to a context.}
\texttt{The context includes information relevant to the nature or generation of the response.}
\texttt{You will assess the quality of the response subject to an evaluation criteria.}

\texttt{\#\#\#Context:}
\texttt{{context\_variables}}

\texttt{\#\#\#Response:}
\texttt{{response}}

\texttt{\#\#\#Evaluation criteria:}
\texttt{{criteria}}
\texttt{{options}}

\texttt{Briefly assess the quality of the response subject to the evaluation criteria.}
\texttt{Focus on the evaluation criteria during assessment, do not provide a general assessment.} \texttt{[/INST]}

\texttt{Assessment: }

\subsubsection{Summarization Prompt}
\noindent\rule{\textwidth}{0pt}
\texttt{<s> [INST]}
\texttt{Summarize the following assessment while maintaining the pertinent details.}
\texttt{Assessment: {assessment}}
\texttt{[/INST]}

\texttt{Assessment Summary:}
\subsubsection{Answer Selection Prompt}
\noindent\rule{\textwidth}{0pt}
\texttt{</s> [INST] Now consider the evaluation criteria and choose a final answer.}

\texttt{Validate the answer against the assessment.}

\texttt{\#\#\#Evaluation criteria:}
\texttt{{criteria}}
\texttt{{options}}
\texttt{[/INST]}
\texttt{Answer:}

\subsection{llama-3-8b-instruct and llama-3-70b-instruct}
\subsubsection{Assessment Prompt}\noindent\rule{\textwidth}{0pt}
\texttt{<|begin\_of\_text|>}
\texttt{<|start\_header\_id|>system<|end\_header\_id|>}
\texttt{You are an fair and objective evaluator.<|eot\_id|>}\\
\texttt{<|start\_header\_id|>user<|end\_header\_id|>}
\texttt{You are presented with a response generated subject to a context. 
The context includes information relevant to the nature or generation of the response.}
\texttt{You will assess the quality of the response subject to an evaluation criteria.}

\texttt{\#\#\#Context:}
\texttt{\{context\_variables\}}

\texttt{\#\#\#Response:}
\texttt{\{response\}}

\texttt{\#\#\#Evaluation criteria:}
\texttt{{criteria}}
\texttt{{options}}

\texttt{Briefly assess the quality of the response subject to the evaluation criteria.}
\texttt{Focus on the evaluation criteria during assessment, do not provide a general assessment.}

\texttt{Assessment: <|eot\_id|>}

\texttt{<|start\_header\_id|>assistant<|end\_header\_id|>}
\subsubsection{Summarization Prompt}\noindent\rule{\textwidth}{0pt}
\texttt{<|begin\_of\_text|>\\<|start\_header\_id|>system<|end\_header\_id|>}

\texttt{You are a helpful assistant<|eot\_id|>\\<|start\_header\_id|>user<|end\_header\_id|>}

\texttt{Summarize the following assessment while maintaining the pertinent details.}

\texttt{Assessment: \{assessment\}}

\texttt{Assessment Summary:<|eot\_id|>}

\texttt{<|start\_header\_id|>assistant<|end\_header\_id|>}

\subsubsection{Answer Selection Prompt}\noindent\rule{\textwidth}{0pt}
\texttt{<|eot\_id|>}
\texttt{<|start\_header\_id|>user<|end\_header\_id|>}

\texttt{Now consider the evaluation criteria and choose a final answer.
Validate the answer against the assessment.}

\texttt{\#\#\#Evaluation criteria:}

\texttt{{criteria}}

\texttt{{options}}

\texttt{Answer:}

\texttt{<|eot\_id|><|start\_header\_id|>}

\texttt{assistant}

\texttt{<|end\_header\_id|>}

\subsection{Prometheus-8x7b-v2}
%\noindent\rule{\textwidth}{0pt}
\texttt{\#\#\#Task Description:}

\texttt{A context that includes information relevant to the nature or generation of the response, a response to evaluate, and a score rubric representing an evaluation criteria are given.}

\texttt{1. Write a detailed feedback that assess the quality of the response strictly based on the given score rubric, not evaluating in general.}

\texttt{2. After writing a feedback, choose a score from the score rubric. Choose one of: \{score\_instructions\}.}

\texttt{3. The output format should look as follows: "Feedback: (write a feedback for criteria) [RESULT] (Choose one of: \{score\_instructions\})"}

\texttt{4. Please do not generate any other opening, closing, or explanations.}

\texttt{\#\#\#Context:}
\texttt{\{context\}}

\texttt{\#\#\#Response to evaluate:}
\texttt{\{response\}}

\texttt{\#\#\#Score Rubrics:}
\texttt{[\{criteria\}]}
\texttt{\{score\_rubric\}}

\texttt{\#\#\#Feedback:}
\section{Pairwise Assessment Prompts for EvalAssist AI Evaluators}
\subsection{Mixtral-8x7b-instruct-v01}
\subsubsection{Assessment Prompt}\noindent\rule{\textwidth}{0pt}
\texttt{<s> [INST] You are provided a pair of responses (Response \{option\_a\ and Response \{option\_b\}) generated subject to a context. }
\texttt{You will choose the better quality response subject to the quality criteria. }

\texttt{This is the context:}
\texttt{\{context\_variables\}}

\texttt{This is the quality criteria:}\\
\texttt{\{criteria\_name\}}
\texttt{\{criteria\_description\}}

\texttt{Response \{option\_a\}:}
\texttt{\{response\_a\}}

\texttt{Response \{option\_b\}:}
\texttt{\{response\_b\}}

\texttt{Keeping the quality criteria in mind, briefly assess which response is better.}
\texttt{Focus on the quality criteria during assessment, do not provide a general assessment.[/INST]}
\texttt{Assessment:}

\subsubsection{Summarization Prompt}\noindent\rule{\textwidth}{0pt}
\texttt{<s> [INST]}
\texttt{Summarize the following assessment while maintaining the pertinent details.}
\texttt{Assessment: \{assessment\}[/INST]}
\texttt{Assessment Summary:}

\subsubsection{Answer Selection Prompt}\noindent\rule{\textwidth}{0pt}

\texttt{</s> [INST] Now considering the quality criteria, which response is better quality?}
\texttt{Validate the answer against the assessment.}
\texttt{\{options [/INST]}
\texttt{Answer:}

\subsection{llama-3-8b-instruct and llama-3-70b-instruct}

\subsubsection{Assessment Prompt}\noindent\rule{\textwidth}{0pt}
\texttt{<|begin\_of\_text|>}
\texttt{<|start\_header\_id|>system<|end\_header\_id|>}
\texttt{You are an fair and objective evaluator.<|eot\_id|>}\\
\texttt{<|start\_header\_id|>user<|end\_header\_id|>}
\texttt{You are provided a pair of responses (Response \{option\_a\} and Response \{option\_b\}) generated subject to a context. }
\texttt{You will choose the better quality response subject to the quality criteria.}

\texttt{This is the context:}
\texttt{\{context\_variables\}}

\texttt{This is the quality criteria:}\\
\texttt{\{criteria\_name\}}
\texttt{\{criteria\_description\}}

\texttt{Response \{option\_a\}:}
\texttt{\{response\_a\}}

\texttt{Response \{option\_b\}:}
\texttt{\{response\_b\}}

\texttt{Keeping the quality criteria in mind, briefly assess which response is better.}
\texttt{Focus on the quality criteria during assessment, do not provide a general assessment.}

\texttt{<|eot\_id|>}

\texttt{<|start\_header\_id|>assistant<|end\_header\_id|>}

\texttt{Assessment:}

\subsubsection{Summarization Prompt}\noindent\rule{\textwidth}{0pt}
\texttt{<|begin\_of\_text|><|start\_header\_id|>}

\texttt{system<|end\_header\_id|>}

\texttt{<|eot\_id|><|start\_header\_id|>user<|end\_header\_id|>}

\texttt{Summarize the following assessment while maintaining the pertinent details.}
\texttt{Assessment: \{assessment\}}

\texttt{Assessment Summary:<|eot\_id|>}

\texttt{<|start\_header\_id|>assistant<|end\_header\_id|>}

\texttt{Assessment Summary:}
    
\subsubsection{Answer Selection Prompt}\noindent\rule{\textwidth}{0pt}
\texttt{<|eot\_id|>}
\texttt{<|start\_header\_id|>user<|end\_header\_id|>}

\texttt{Now consider the evaluation criteria and choose a final answer.
which response is better quality? Validate the answer against the assessment.}

\texttt{{options}}

\texttt{<|eot\_id|><|start\_header\_id|>assistant<|end\_header\_id|>}

\texttt{Answer:}

\subsection{Prometheus-8x7b-v2}

\texttt{\#\#\#Task Description:}
\texttt{A context that includes information relevant to the nature or generation of the response, a response to evaluate, and a score rubric representing a evaluation criteria are given.}

\texttt{1. Write a detailed feedback that assess the quality of two responses strictly based on the given score rubric, not evaluating in general.}

\texttt{2. After writing a feedback, choose a better response between Response \{option\_a\} and Response \{option\_b\}. You should refer to the score rubric.}

\texttt{3. The output format should look as follows: "Feedback: (write a feedback for criteria) [RESULT] (\{option\_a\} or \{option\_b\})”}

\texttt{4. Please do not generate any other opening, closing, and explanations.}

\texttt{\#\#\#Context:}
\texttt{\{context\_variables\}}

\texttt{\#\#\#Response \{option\_a\}:}
\texttt{\{response\_1\}}

\texttt{\#\#\#Response \{option\_b\}:}
\texttt{\{response\_1\}}

\texttt{\#\#\#Score Rubric:}
\texttt{\{rubric\}}

\texttt{\#\#\#Feedback: }

%\section{Figures}
%\begin{figure*}
 %   \centering
  %  \includegraphics[width=1\linewidth]{figures/agg100.png}
   % \caption{Aggregated changes in evaluation criteria by task and judgment strategy.}
   % \label{fig:aggregrated}
   % \end{figure*}

\end{document}